\documentclass[useAMS,usenatbib]{mn2e}

\usepackage{float}
\usepackage{color}
\usepackage{lscape,graphicx}
\usepackage{amsmath}
\usepackage{journal_shortcuts}
\usepackage{amssymb}
\usepackage{multirow}
\usepackage{afterpage}
\usepackage{ulem}
\usepackage{threeparttable}
\usepackage{subfig}
\def\ga{{\rm\thinspace gauss}}

\def\h0{\hbox{{\rm H}$^0$}}
\DeclareMathAlphabet{\vib}{OML}{cmm}{m}{it}
\newcommand*{\satellite}[1]{\textit{#1}}

\newcommand*{\rosat}{\satellite{ROSAT}}
\newcommand*{\xmm}{\satellite{XMM-Newton}}
\newcommand*{\chandra}{\satellite{Chandra}}

\newcommand*{\gmrt}{\satellite{GMRT}}
\newcommand*{\vla}{\satellite{VLA}}

\newcommand{\lsim}{\mathrel{\hbox{\rlap{\lower.55ex\hbox{$\sim$}} \kern-.3em \raise.4ex \hbox{$<$}}}}
\newcommand{\gsim}{\mathrel{\hbox{\rlap{\lower.55ex\hbox{$\sim$}} \kern-.3em \raise.4ex \hbox{$>$}}}}

\title[No shock across part of a relic in ZwCl 2341?]{No Shock Across Part of a Radio Relic in the Merging Galaxy Cluster ZwCl 2341.1+0000?}
\author[G.~A.~Ogrean et al.]{G.~A.~Ogrean$^{1}$\thanks{E-mail:
gogrean@hs.uni-hamburg.de}, M. Br\"uggen$^{1}$, R.~J.~van~Weeren$^{2}$, A.~Burgmeier$^{3}$, A. Simionescu$^{4}$\\
$^{1}$Hamburger Sternwarte, Gojenbergsweg 112, 21029 Hamburg, Germany\\
$^{2}$Harvard-Smithsonian Center for Astrophysics, 60 Garden Street, Cambridge, MA 02138, USA\\
$^{3}$Deutsches Elektronen-Synchrotron (DESY), Notkestra{\ss}e 85, 22607 Hamburg, Germany\\
$^{4}$Japan Aerospace Exploration Agency, 7-44-1 Jindaiji Higashimachi Chofu, Tokyo 182-8522 Japan}

\begin{document}

\date{Accepted xxx xxxx xx. Received xxx xxxx xx; in original form xxx xxxx xx}

\pagerange{\pageref{firstpage}--\pageref{lastpage}} \pubyear{2014}

\maketitle

\label{firstpage}

\begin{abstract}

\noindent The galaxy cluster ZwCl~2341.1+0000 is a merging system at $z=0.27$, which hosts two radio relics and a central, faint, filamentary radio structure. The two radio relics have unusually flat integrated spectral indices of $-0.49\pm 0.18$ and $-0.76\pm 0.17$, values that cannot be easily reconciled with the theory of standard diffusive shock acceleration of thermal particles at weak merger shocks. We present imaging results from \xmm\ and \chandra\ observations of the cluster, aimed to detect and characterise density discontinuities in the ICM. As expected, we detect a density discontinuity near each of the radio relics. However, if these discontinuities are the shock fronts that fueled the radio emission, then their Mach numbers are surprisingly low, both $\leq 2$. We studied the aperture of the density discontinuities, and found that while the NW discontinuity spans the whole length of the NW radio relic, the arc spanned by the SE discontinuity is shorter than the arc spanned by the SE relic. This startling result is in apparent contradiction with our current understanding of the origin of radio relics. Deeper X-ray data are required to confirm our results and to determine the nature of the density discontinuities.

\end{abstract}

\begin{keywords}
 galaxies: clusters: individual: ZwCl $2341.1+0000$ -- X-rays: galaxies: clusters -- shock waves
\end{keywords}

\section{Introduction}
\label{sec:intro}

Galaxy clusters grow by merging with smaller clusters and groups, and by accreting gas from the intergalactic medium. Part of the kinetic energy associated with cluster mergers is dissipated in weak shocks with Mach numbers $\mathcal{M} \approx 2$, which travel outwards through the intracluster medium (ICM). The shocks are capable of accelerating particles to relativistic energies, at which the newly formed cosmic rays (CRs) emit synchrotron radiation at radio frequencies. Indeed, diffuse radio sources known as ``radio relics'' have been found exclusively in the outskirts of several merging clusters, and some of them trace weak shock fronts. Radio relics have low brightness, steep spectra ($\alpha<-1$, with the radio flux at frequency $\nu$ being $\mathcal{F}\propto \nu^{\alpha}$), and sizes ranging from several hundred kpc up to $1-2$ Mpc. Early on, it was proposed that relics are formed when merger-triggered shocks accelerate thermal particles via diffusive shock acceleration \citep[DSA; e.g.,][]{Ensslin1998}.\footnote{Here, we refer to this acceleration model as the standard DSA model.} The predictions of the standard DSA model are that (i) merger shocks are associated with radio relics, (ii) shock fronts trace the outer edges of the radio relics, and (iii) the Mach number of shocks can be independently determined from the X-ray and the radio data.

Recent results, however, point towards a more complex acceleration mechanism. In the galaxy cluster Abell 2146, merger shocks discovered in the cluster outskirts are not associated with radio relics \citep{Russell2011}. In the merging cluster 1RXS~J0603.3+4214, the largest radio relic is spatially offset from the shock front, and there is a significant discrepancy between the X-ray- and radio-derived Mach numbers \citep{vanWeeren2012,Ogrean2013a}. In the cluster CIZA~J2242.8+5301, the Mach numbers derived from the X-ray and radio data for one of the two relics are also discrepant; additionally, there are pairs of consecutive shocks on both sides of the merger axis, and only the outer shocks in each pair are associated with relics \citep{vanWeeren2010,Ogrean2014}. From a theoretical standpoint, the efficiency of particle acceleration at weak shocks is also too low to easily explain the observed brightness of radio relics via standard DSA \citep[e.g.,][]{Kang2007}.

One possible solution to these multiple puzzles is the presence of a pre-existing cosmic ray (CR) electron population, which is re-accelerated by merger shocks. If relics are created when shocks re-accelerate aged CR particle populations (which are no longer visible in the radio), then the acceleration efficiency required to produce the observed radio brightness is diminished. Furthermore, no radio emission would be produced in the absence of pre-existing CRs, which explains the occasional offset between relics and shocks, and even the complete absence of relics at some shock fronts. Other possibile explanations also exist, including a highly inhomogeneous magnetic field strength and projection effects. 

While some merging clusters host one \citep[e.g., Abell~521,][]{Ferrari2006}, two \citep[e.g., CIZA~J2242.8+5301][]{vanWeeren2010}, or more radio relics \citep[e.g., 1RXS~J0603.3+4214][]{vanWeeren2012}, others host more central Mpc-scale diffuse radio structures that roughly follow the X-ray surface brightness; these central radio sources are known as ``radio halos''. Halos also have steep spectra, but unlike radio relics, they are not clearly polarized. Their origin is a matter of debate. Two alternative scenarios are currently preferred: in-situ re-acceleration by turbulence amplified during cluster mergers \citep[e.g.,][]{Brunetti2001}, and injection of relativistic electrons in the ICM by hadronic collisions \citep[e.g.,][]{BlasiCola1999}. In a few instances, merger shocks were discovered to be spatially coincident with part of the radio halo \citep[e.g.,][]{Markevitch2002,Brown2011}. There are a handful of merging clusters hosting both radio halos and radio relics, but also merging clusters with shocks yet without halos and relics \citep[e.g., Abell~2146,][]{Russell2011}. The diversity in radio structure statistics in different merging galaxy clusters raises important questions related to the origin of diffuse radio emission: If shocks are responsible for the origin of radio relics, why do not all merging clusters host radio relics? Why do some clusters host only relics, only halos, both, or no extended radio structures? Does the presence of a radio halo or of relics depend on the stage of the merger? Does it depend on the merger scenario or on the orientation of the merger axis with respect to our line of sight? Answering these questions is made difficult at the moment by the small number of well-understood merging clusters. A detailed characterization of merging clusters is only possible with deep multiwavelength observations, and often with complementary numerical simulations \citep[e.g.,][]{vanWeeren2011b,Brueggen2012}.

Here, we present results from moderately-deep \xmm\ and \chandra\ observations of the merging galaxy cluster ZwCl~2341.1+0000 ($z=0.27$). Preliminary \chandra\ images revealed a NW-SE elongated cluster, suggesting this direction as the orientation of the merger axis \citep{vanWeeren2009b}. \emph{Giant Metrewave Radio Telescope} (\gmrt) images at 241 and 610 MHz showed that the cluster hosts a double radio relic system, with the two relics located approximately perpendicular to the putative merger axis, one on each side of the cluster. Between 157 and 1400 MHz, the NW and SE relics have very flat integrated spectral indices, of $-0.49\pm 0.18$ and $-0.76\pm 0.17$, respectively {vanWeeren2009b}. Among the $\approx 20$ known radio relics, only one other has a spectral index significantly above $-1$ -- the relic in Abell~2256, with an integrated spectral index of $-0.81\pm 0.03$ \citep{vanWeeren2012b}. ZwCl~2341.1+0000 has also been observed with the \vla\ in D-array configuration at 1.4~GHz. These observations confirmed the two relics and revealed additional 2.2-Mpc central radio emission \citep{Giovannini2010}. The central radio source was classified as a filament by \citet{Giovannini2010}, but alternative explanations for its origin have also been proposed, e.g. a very large radio halo, or the combined radio halos of at least two merging clusters. However, the source is relatively strongly polarized, and its radio power is very low compared to that of other radio halos of similar size \citep{Giovannini2009,Giovannini2010}. 

The main aims of our analysis of \xmm\ and \chandra\ data of ZwCl~2341.1+0000 are to detect and characterize shocks within the ICM, and to compare the properties of the shock(s) with those of the radio structures. The analysis has implications on our understanding of the cluster's merger history and, more generally, of particle acceleration at merger shocks.

This paper is organized as follows: In Section \ref{sec:data-reduction}, we describe the X-ray observations and the data reduction steps. The results of our analysis are presented in Section \ref{sec:analysis}. In Section \ref{sec:discussion}, we summarize the results and discuss the correspondence between the detected discontinuities and the radio structures.

We assume a $\Lambda$CDM cosmology with $H_0=70$ ${\rm km\,\,s^{-1}\,\,Mpc^{-1}}$, $\Omega_{\rm m}=0.3$, and $\Omega_{\rm \Lambda}=0.7$. At the distance of ZwCl~2341.1+0000, 1 arcmin corresponds to approximately 246 kpc. Unless specified otherwise, errors quoted throughout the paper are $1\sigma$ statistical errors.

\section{Data reduction}
\label{sec:data-reduction}

\subsection{Chandra}

ZwCl~2341.1+0000 was observed for 30~ks with \chandra/ACIS-I on October 18, 2006 (ObsID 5786). The observation was carried out in Very Faint (VF) mode. We reduced the dataset with CIAO v4.6. Periods of high background were detected with \emph{lc\_clean} using the S3 chip and the energy band $2.5-7$ keV, and they were removed from the data. After the removal of the background flares, the cleaned event file had an exposure time of 25~ks. Point sources were detected with the task \emph{wavdetect} using wavelet radii of $1, 2, 4, 8$, and 16 pixels, confirmed by eye, and removed from the event file.

The instrumental background was modelled using the D-period stowed background event file, which was processed following the ACIS VF background cleaning method to filter out events flagged as potential background events. The background was renormalized to have the same hard band ($10-12$~keV) count rate as the cluster observation.

\begin{figure*}
	\centering
	\includegraphics[width=0.495\textwidth, clip=true, keepaspectratio=true]{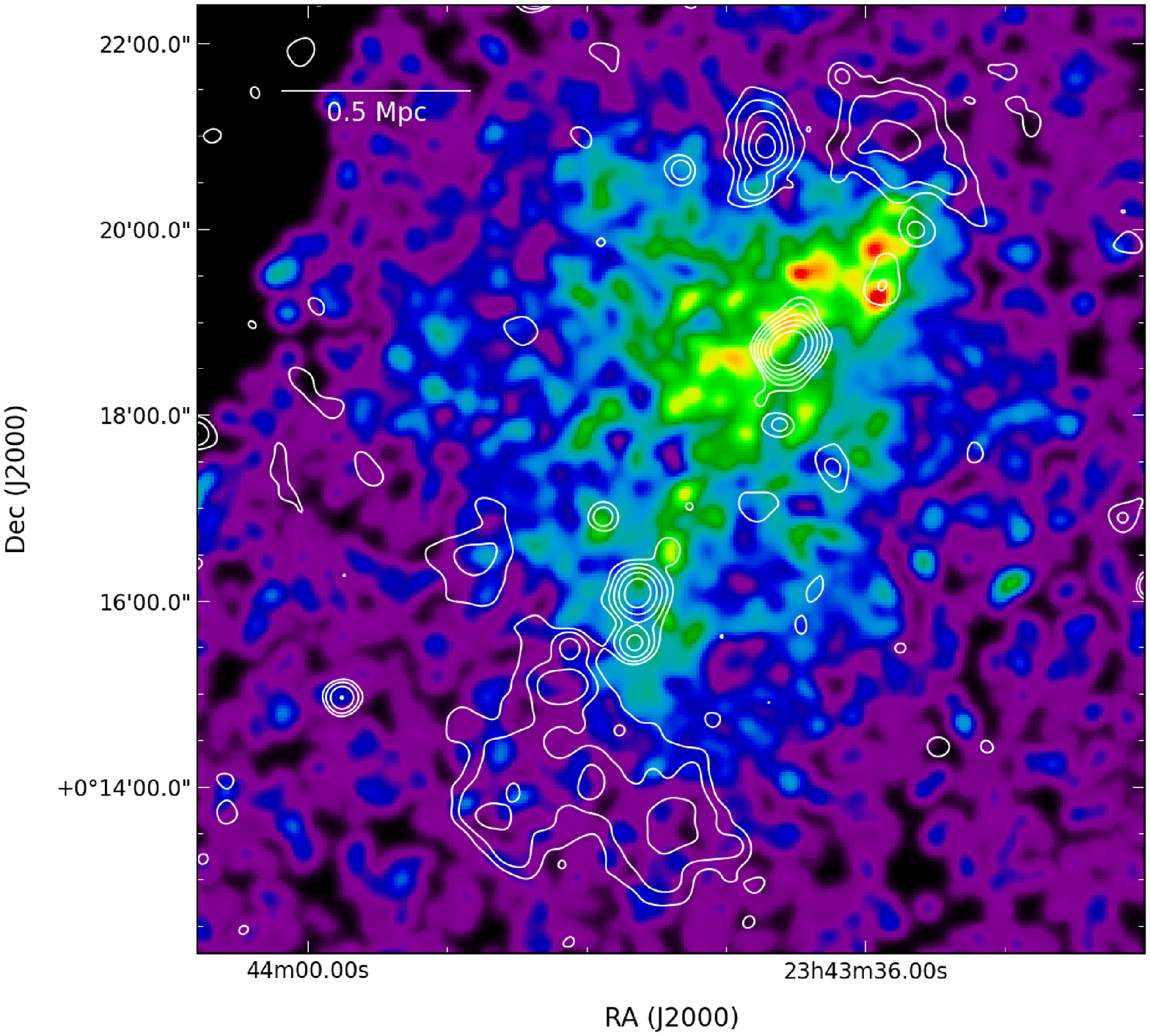}
	\includegraphics[width=0.495\textwidth, clip=true, keepaspectratio=true]{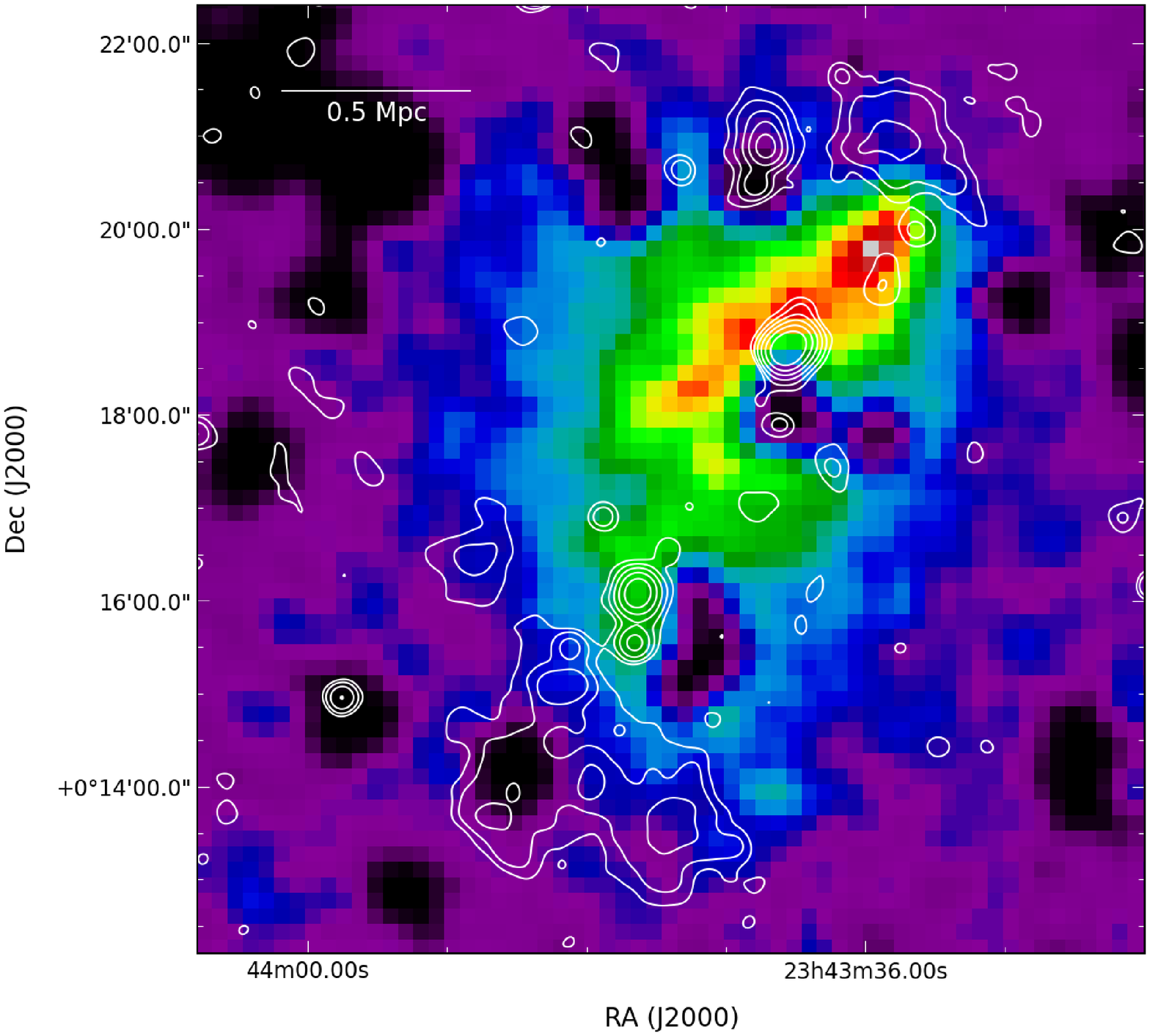}
	\caption{\emph{Left:} \chandra\ surface brightness map in the energy band $0.5-3$ keV. The image was binned by 4, exposure-corrected, and smoothed with a Gaussian kernel of size $3\times 3$ pixels (1 pixel $= 1.97$ arcsec). Point sources were subtracted, and the gaps were filled by sampling the Poisson distribution whose mean is that of the pixels surrounding the corresponding excluded point sources. Naturally, this highly-processed image is only a ``pretty image'', and has not been used in the scientific interpretation. \emph{Right:} \xmm\ surface brightness map in the energy band $0.5-4$ keV. The image was binned by 2, exposure-corrected, instrumental background-subtracted, and smoothed with a Gaussian kernel of size $3\times 3$ pixels (1 pixel $= 10$ arcsec). Gaps in the image indicate subtracted point sources.}
	\label{fig:prettypictures}
\end{figure*}

\subsection{XMM-Newton}

ZwCl~2341.1+0000 was observed on June 12, 2005, with the European Photon Imaging Camera (EPIC) on board of \xmm\ (ObsID 0211280101). The observation used the thin filter, and had a total exposure time of 42.6 ks. We analyzed the dataset with the \xmm\ Extended Source Analysis Software \citep[{\sc esas}, e.g.][]{KuntzSnowden2008,Snowden2008} integrated in the Scientific Analysis System ({\sc sas}) v12.0.1, and the latest calibration files as of December 15, 2012. 

Raw MOS and pn event files were generated from the Observation Data Files (ODF) using the \emph{emchain} and \emph{epchain} routines. Each of the event files was examined for soft proton flares\footnote{Time-variable component affecting the exposed detector pixels, created when protons with energies less than a few hundred keV interact with the telescope and are funneled towards the detectors.}; time periods with count rates outside a $3\sigma$ interval around the peaks of the Gaussians fitted to the three EPIC light curves were excluded from further analysis. The cleaned event files have exposure times of 30.0 (MOS1 and MOS2), and 20.4 ks (pn). Point sources were selected from the 3XMM-DR4 Serendipitous Source Catalogue, confirmed visually, and excluded from the analysis. A few additional point sources and detector artefacts have also been removed. Out-of-time events, recorded during the readout time of the CCDs, have been excluded as well.

The instrumental background of each detector was modelled using data from outside the field of view (FOV; the unexposed corners of each detector), and from filter-wheel closed (FWC) observations with similar hardness ratios and count rates as the cluster observation. The background was then renormalized to have the same count rate in the energy band $10-12$~keV as that measured in the cluster observation in the same energy band.

\section{Analysis}
\label{sec:analysis}

\subsection{Imaging analysis}
\label{sec:imaging}

In Figure \ref{fig:prettypictures} we show the \emph{Chandra} and \emph{XMM-Newton} surface brightness maps. The cluster is clearly a merger elongated along the NW-SE direction, which supports the interpretation of this direction as the merger axis. Off the merger axis, excess X-ray emission is present to the E-NE, perpendicular to the merger axis. The surface brightness decreases rapidly ahead of the NW bullet-like subcluster and SE of the tail-like SE structure, suggesting possible density discontinuities near the two radio relics.

To model the surface brigthness profiles across the putative density discontinuities, we created $0.5-4$~keV \xmm\ and $0.5-3$~keV \chandra\ surface brightness profiles in elliptical sectors across the radio relics. The energy ranges were chosen to maximize the signal-to-noise. The \xmm\ and \chandra\ surface brightness profiles were fitted in parallel, under the assumption that they are both described by the same underlying density model. The models fitted to the \xmm\ profiles were corrected for point spread function (PSF) effects using a King model with core radius of $4.7$ arcsec, and a slope of $1.5$ (the average core radius and slope calculated from the best-fit MOS and pn on-axis values at 1.5~keV).

We modelled the $0.5-4$~keV \xmm\ and $0.5-3$~keV \chandra\ non-X-ray background (NXB) in each elliptical sector using instrumental background event files. The instrumental background surface brightness profiles were not subtracted from the cluster profiles, but rather added to the models describing the cluster emission and the sky background. This approach preserves the Poissonian structure of the count distribution, and allows the use of Cash statistics \citep{Cash1979} when fitting the surface brightness profiles. The Cash statistical method is more accurate than the $\chi^2$ method when fitting data with a low number of counts per bin, and hence it is more adequate for our rather short archival observations.

\begin{figure}
	\centering
	\includegraphics[width=\columnwidth, clip=true, keepaspectratio=true]{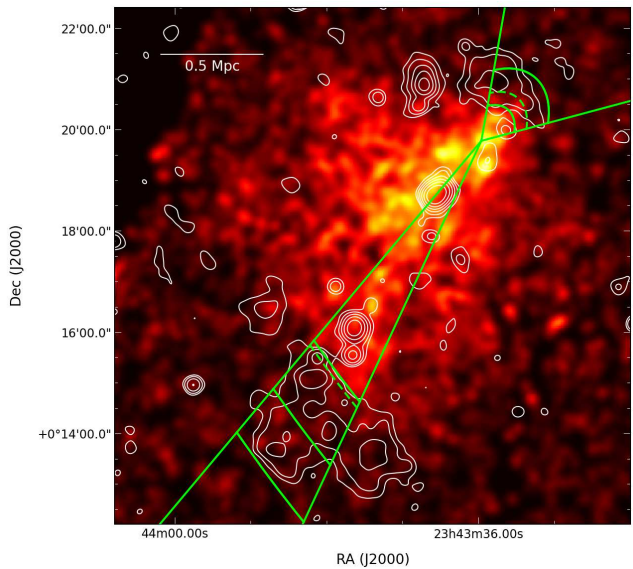}
	\caption{Elliptical sectors used for extracting the \chandra\ and \xmm\ surface brightness profiles, overlaid on the \chandra\ map shown in Figure \ref{fig:prettypictures}. Solid-line elliptical arcs mark the inner and outer limits of the relics, and are drawn at the same distances as the red lines in the surface brightness profiles (see Figs. \ref{fig:s-best-fits} and \ref{fig:n-best-fits}). Dashed lines mark the distances of the density discontinuities detected in each profile.\label{fig:sectors}}
\end{figure}

\begin{table*}
\caption{Best-fit parameters of the models fitted in parallel to the \xmm\ and \chandra\ surface brightness profiles across the SE relic. The fits were done using Cash statistics, hence we do not provide a goodness of the fit.}
\label{tab:s-best-fits}
\centering
\begin{threeparttable}
\begin{tabular}{cccccc}
 \multicolumn{6}{c}{{\sc SE Sector}} \\
 \multicolumn{6}{c}{{\sc $\beta$-Model}} \\
 \hline
     	\multicolumn{2}{c}{$\beta$\tnote{$\dagger$}} & \multicolumn{2}{c}{$r_c$ (arcmin)} & ${\rm SB}_{0,\,\,{\rm XMM}}$ (cts s$^{-1}$ arcmin$^{-2}$) & ${\rm SB}_{0,\,\,{\rm Chandra}}$  (cts s$^{-1}$ arcmin$^{-2}$) \\ 
 \hline
	\multicolumn{2}{c}{$0.85$} & \multicolumn{2}{c}{$7.82_{-0.37}^{+0.38}$} & $1.27_{-0.12}^{+0.13} \times 10^{-2}$ & $1.88_{-0.21}^{+0.25} \times 10^{-2}$ \\
 \hline
    & & & & \\
 \multicolumn{6}{c}{{\sc Broken Power-Law Model}} \\
 \hline
     	$\alpha_1$ & $\alpha_2$ & $r_{\rm d}$ (arcmin) & $C$ &  ${\rm SB}_{0,\,\,{\rm XMM}}$ (cts s$^{-1}$ arcmin$^{-2}$) & ${\rm SB}_{0,\,\,{\rm Chandra}}$  (cts s$^{-1}$ arcmin$^{-2}$) \\ 
 \hline
	$-0.40_{-0.21}^{+0.20}$ & $3.76_{-0.69}^{+0.83}$ & $10.59_{-0.15}^{+0.10}$ & $1.62_{-0.27}^{+0.32}$ & $5.88_{-0.60}^{+0.66} \times 10^{-4}$ & $8.41_{-0.89}^{+0.97} \times 10^{-4}$ \\
\hline
\end{tabular}
	\begin{tablenotes}
		\item[$\dagger$] fixed parameter
	\end{tablenotes}
   \end{threeparttable}
\end{table*}

\begin{figure*}
 	\begin{center}
  		\includegraphics[width=0.495\textwidth,keepaspectratio=true,clip=true]{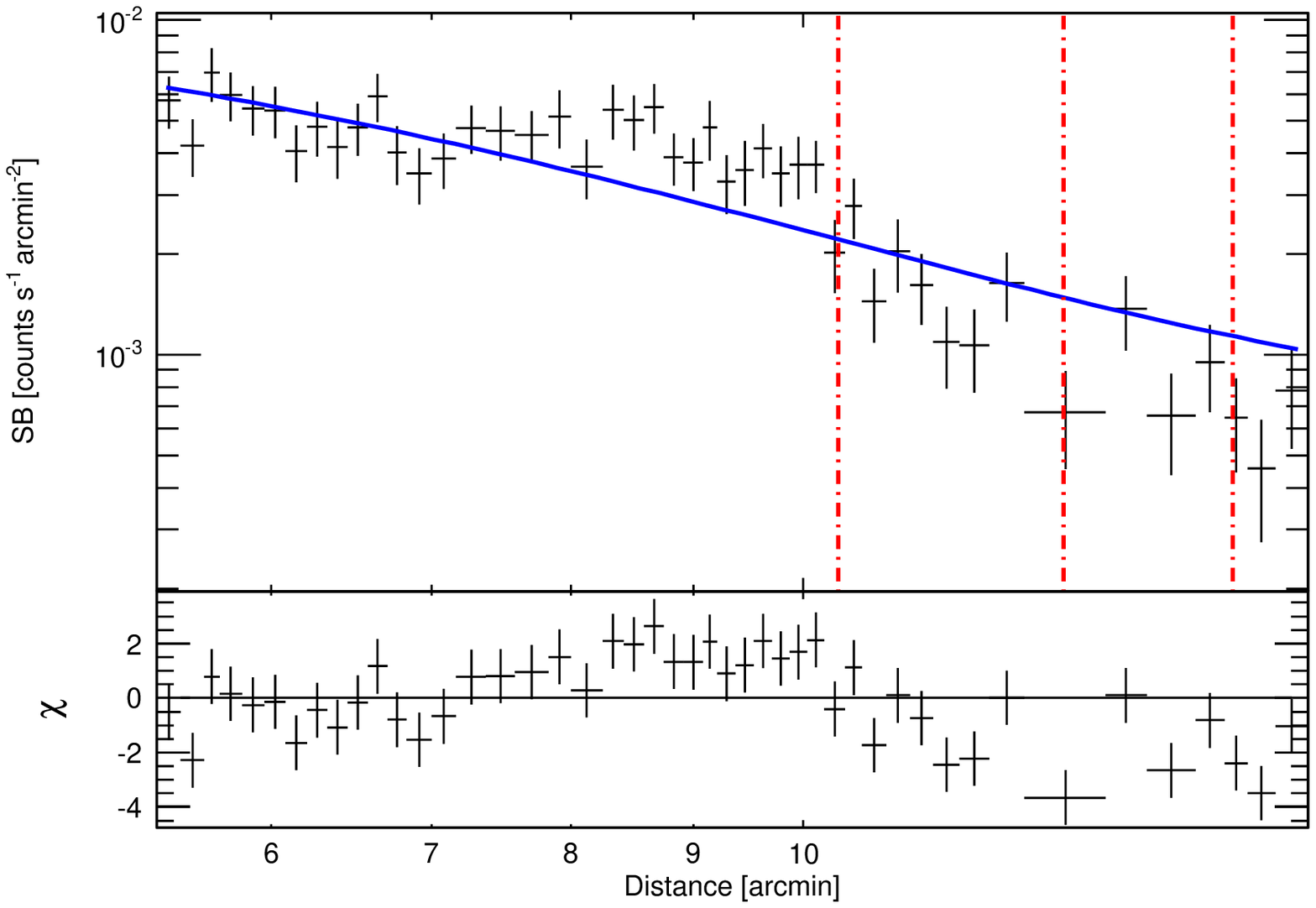}
		\includegraphics[width=0.495\textwidth,keepaspectratio=true,clip=true]{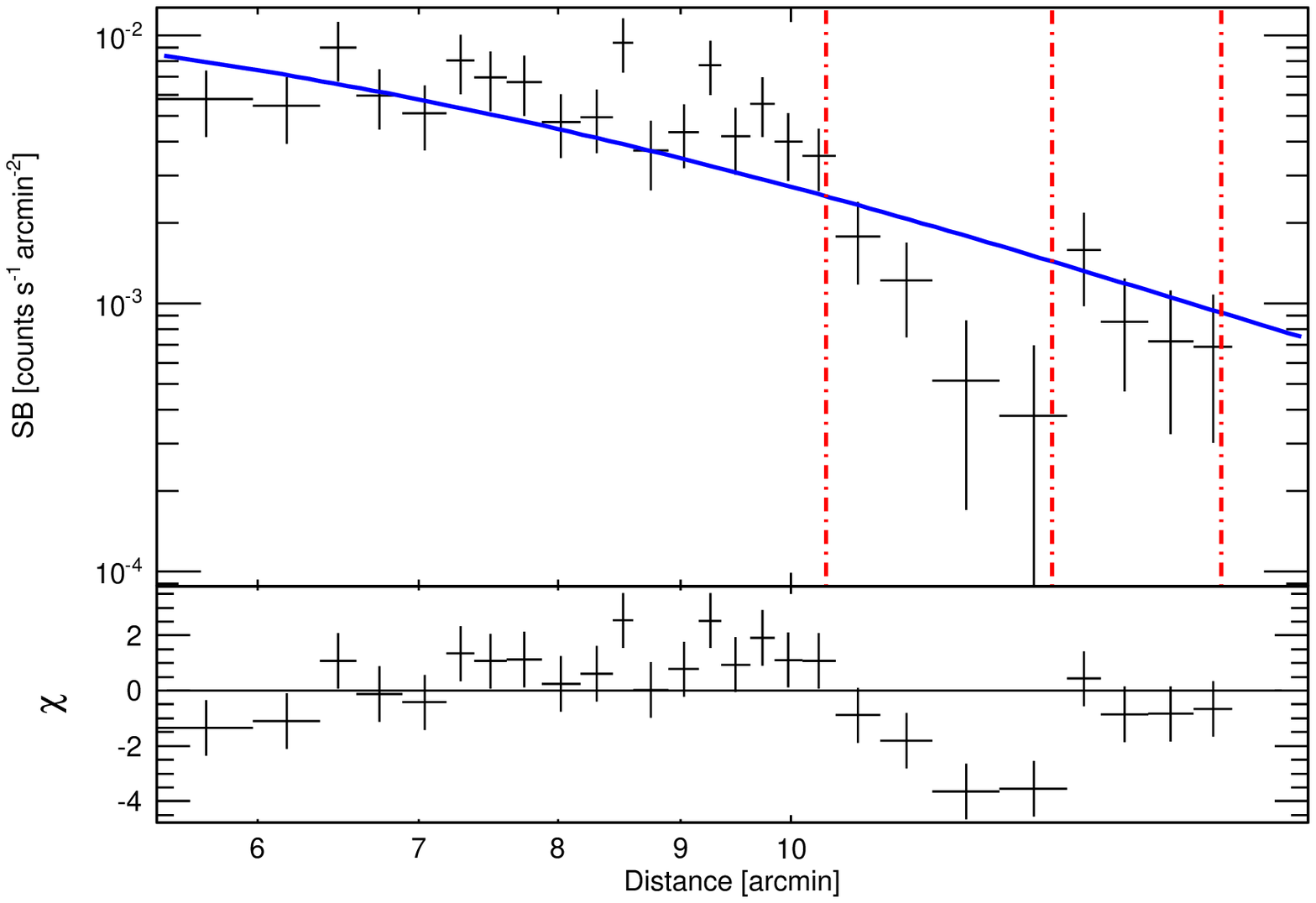}
		\includegraphics[width=0.495\textwidth,keepaspectratio=true,clip=true]{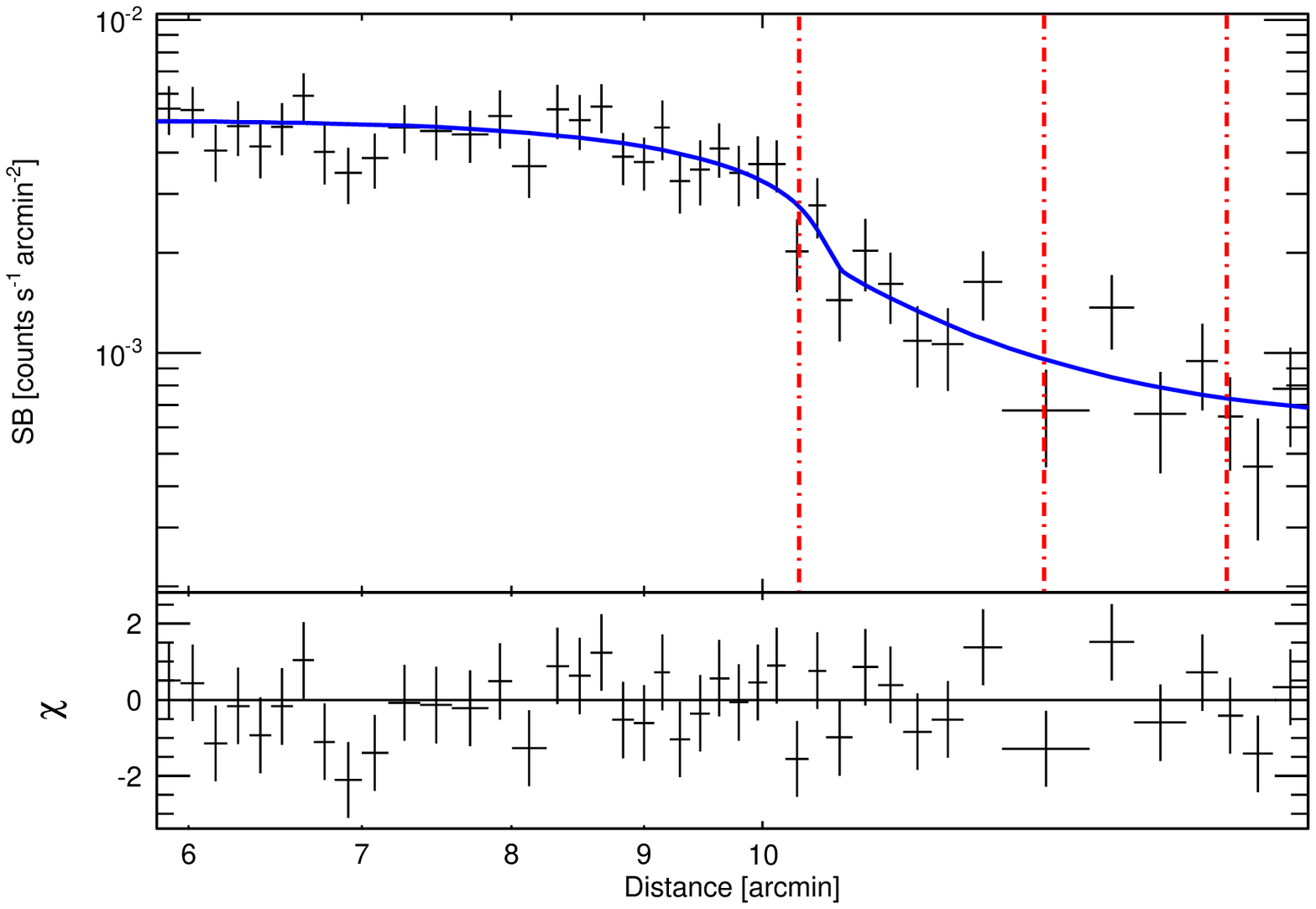}
		\includegraphics[width=0.495\textwidth,keepaspectratio=true,clip=true]{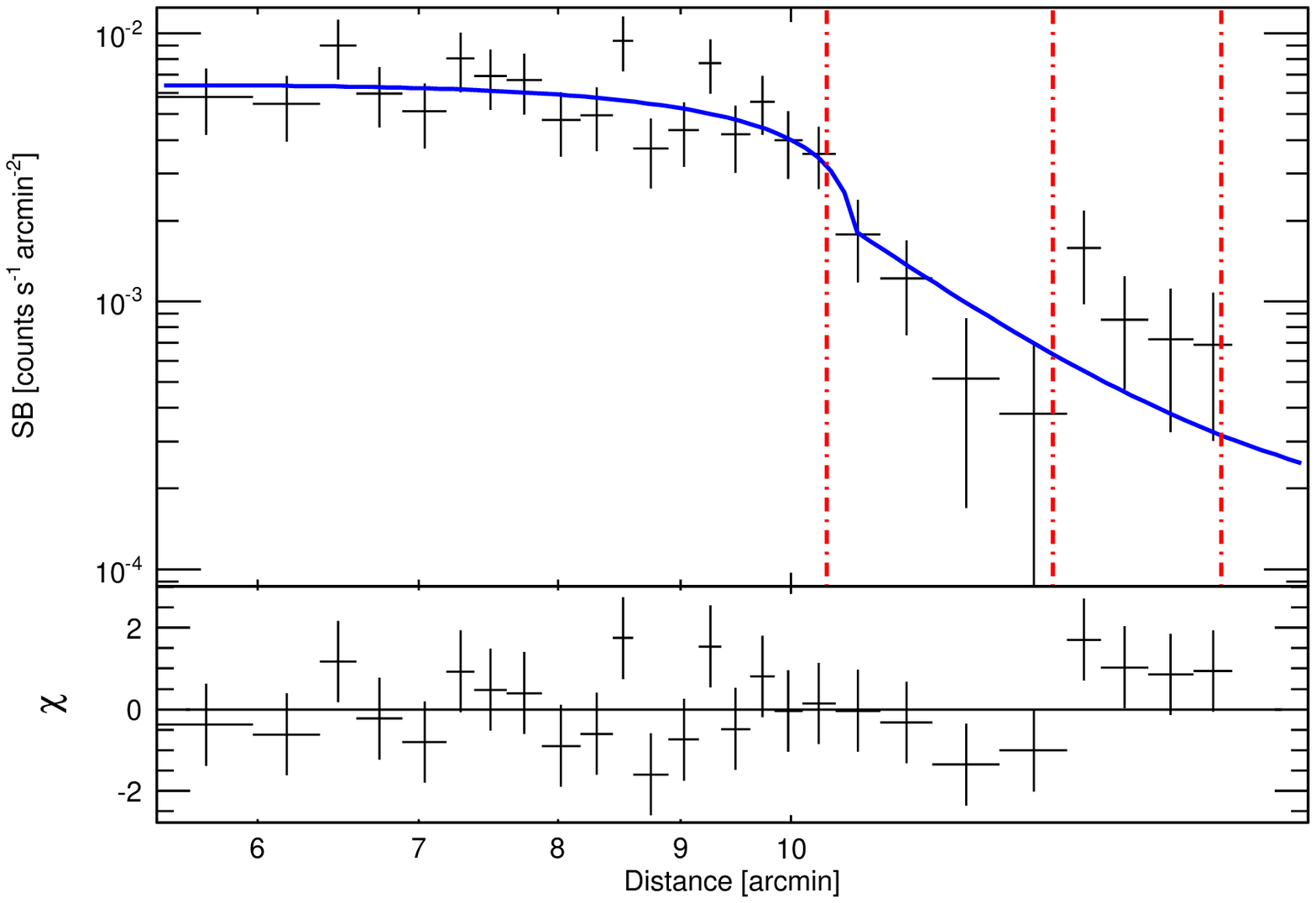}
 	\end{center}
	\caption{Best-fit models, overplotted onto the \xmm\ and \chandra\ surface brightness profiles across the SE relic. Top panels show the $\beta$-model, while the bottom panels show the broken power-law model. The parameters of the models are presented in Table~\ref{tab:s-best-fits}. Red lines correspond to the different edges of the relic (see Figure \ref{fig:sectors}), and are drawn at distances of $10.35$, $12.86$, and $15.12$ arcmin. \xmm\ profiles are in the left panels, and \chandra\ in the right panels. All profiles were regrouped for visual clarity, but the models were actually calculated with a small binning of $\approx 5$ counts/bin for both \xmm\ and \chandra.}
	 \label{fig:s-best-fits}
\end{figure*}

\subsubsection{SE density discontinuity}

The sector across the SE radio relic is shown in Figure \ref{fig:sectors}. The \xmm\ and \chandra\ surface brightness profiles were created in elliptical sectors of opening angles 15 degrees, and were binned to a minimum of 5 counts/bin. The sky background levels were determined by fitting a constant to the outer bins of each of the surface brightness profiles, in the radius range $18-24$ arcmin. The best-fit values are $6.12_{-0.54}^{+0.55}\times 10^{-4}$ and $1.45_{-0.58}^{+0.61} \times 10^{-4}$ cts/s/arcmin$^2$ for \xmm\ and \chandra, respectively.

We first attempted to model the surface brightness profiles in the radius range $5-16$ arcmin with a $\beta$-model:
\begin{eqnarray}
	{\rm SB}(r) = {\rm SB}_0 \, \left[ 1+\left( \frac{r}{r_c} \right)^2 \right]^{-3\beta+0.5}, \nonumber
\end{eqnarray}
where ${\rm SB}(r)$ is the surface brightness at radius $r$, $SB_0$ is the central surface brightness, and $r_c$ is the core radius. The $\beta$-parameter was fixed to $0.85$, the value calculated by \citet{Eckert2012} in the outskirts of non-cool core clusters from the average profile of 17 clusters observed with \rosat/PSPC. The core radii were linked for the \xmm\ and \chandra\ models, but otherwise left free in the fit. The central surface brightness values were free and distinct for the two models. The X-ray background was kept fixed to the values determined in the outer bins of each profile -- $6.12_{-0.54}^{+0.55}\times 10^{-4}$ and $1.45_{-0.58}^{+0.61} \times 10^{-4}$ cts/s/arcmin$^2$ for \xmm\ and \chandra, respectively. The $\beta$-model fits are shown in Figure \ref{fig:s-best-fits}, and the best-fit parameters are summarized in Table \ref{tab:s-best-fits}. The best-fit models clearly deviate from the data, proving that the outskirts of ZwCl~2341.1+0000 are not in hydrostatic equilibrium. It appears that the primary reason for the poor fits is the flatness of the X-ray profiles out to around 9 arcmin, followed by the sudden decrease in brightness in the range $9-11$ arcmin.

The shapes of the profiles suggest that they would be modelled better by a broken power-law model. Therefore, we employed a broken power-law density model of the form:
\begin{eqnarray}
	n_1(r) & = & C\, n_0\, \left(\frac{r}{r_{\rm d}}\right)^{-\alpha}\,, \quad {\rm for} \; r\le r_{\rm d}\,, \nonumber \\
	n_2(r) & = & n_0\, \left(\frac{r}{r_{\rm d}}\right)^{-\beta}\,, \quad {\rm for} \; r>r_{\rm d}\,, \nonumber
\end{eqnarray}
where $n_{\rm i}$ is the number density, $C$ is the density compression, $n_0$ is the density normalization, $r$ is the radius from the centre of the sector, $r_{\rm d}$ is the radius at the location of the density discontinuity, and the indices $1$ and $2$ correspond, respectively, to the putative post-shock and pre-shock regions. We then modeled the surface brightness by projecting this density model along the line of sight, under the assumption that the cluster ICM geometry is described by a prolate spheroid with two of its unequal axes in the plane of the sky\footnote{The prolate spheroid is shaped like a rugby ball with semi-axes ($R_1$, $R_2$, $R_3$) = ($b, a, a$), with $a \ge b$, and may be rotated in the plane of the sky by an angle $\theta$ with respect to the $(x,y)$ coordinate system. The semi-axis $R_3$ of the spheroid is aligned with the line of sight.} \citep{Owers2009}; we also added to the model a constant that describes the X-ray background level. 

\begin{table*}
\caption{Best-fit parameters of the broken power-law models refitted in parallel to the surface brightness profiles across the SE relic after simultaneously varying the \xmm\ and \chandra\ sky background levels by $1\sigma$. The fits were done using Cash statistics, hence we do not provide a goodness of the fit.}
\label{tab:s-robustness}
\centering
\begin{tabular}{cccccc}
 \multicolumn{6}{c}{{\sc SE Sector}} \\
 \multicolumn{6}{c}{{\sc Low Background Level}} \\
 \hline
     	$\alpha_1$ & $\alpha_2$ & $r_{\rm d}$ (arcmin) & $C$ &  ${\rm SB}_{0,\,\,{\rm XMM}}$ (cts s$^{-1}$ arcmin$^{-2}$) & ${\rm SB}_{0,\,\,{\rm Chandra}}$  (cts s$^{-1}$ arcmin$^{-2}$) \\ 
 \hline
	$-0.41_{-0.21}^{+0.20}$ & $3.44_{-0.60}^{+0.69}$ & $10.59_{-0.13}^{+0.10}$ & $1.65_{-0.25}^{+0.31}$ & $5.92_{-0.61}^{+0.65} \times 10^{-4}$ & $8.43_{-0.89}^{+0.96} \times 10^{-4}$ \\
\hline
    & & & & \\
 \multicolumn{6}{c}{{\sc High Background Level}} \\
 \hline
     	$\alpha_1$ & $\alpha_2$ & $r_{\rm d}$ (arcmin) & $C$ &  ${\rm SB}_{0,\,\,{\rm XMM}}$ (cts s$^{-1}$ arcmin$^{-2}$) & ${\rm SB}_{0,\,\,{\rm Chandra}}$  (cts s$^{-1}$ arcmin$^{-2}$) \\ 
 \hline
	$-0.41_{-0.21}^{+0.20}$ & $4.20_{-0.80}^{+1.03}$ & $10.59_{-0.18}^{+0.10}$ & $1.57_{-0.29}^{+0.33}$ & $5.84_{-0.60}^{+0.66} \times 10^{-4}$ & $8.39_{-0.89}^{+0.98} \times 10^{-4}$ \\
\hline
\end{tabular}
\end{table*}

All the parameters, with the exception of the sky background levels, were free in the fit. The sky background levels were fixed to the best-fit values determined from the outer bins of the corresponding profiles. The broken power-law model was simultaneously fitted to the \xmm\ and \chandra\ data in the radial range $5-16$ arcmin, using Cash statistics. The density compression, power-law indices, and jump radius were linked for \xmm\ and \chandra, so that the same physical model is used to describe both datasets. The best-fit parameters are presented in Table \ref{tab:s-best-fits}, and the model is shown in Figure \ref{fig:s-best-fits}. The best-fit density compression is $1.62_{-0.27}^{+0.32}$. If the discontinuity is a shock front, then the density compression is related with the shock Mach number, $\mathcal{M}$, via:
\begin{eqnarray}
	\frac{1}{C} = \frac{3}{4}\,\frac{1}{\mathcal{M}^2} + \frac{1}{4}, \nonumber
\end{eqnarray}
and hence $1 \le C<4$. A density compression $C=1.62_{-0.27}^{+0.32}$ corresponds to a putative shock with $\mathcal{M} = 1.43_{-0.20}^{+0.23}$.

To test the robustness of our results, we varied the sky background levels within their $1\sigma$ statistical errors. The sky background levels were simultaneously decreased/increased, and the broken power-law density model was refitted in parallel to the \xmm\ and \chandra\ profiles. The best-fit parameters with the lower and higher sky background levels are summarized in Table \ref{tab:s-robustness}. The density compression remains significantly $>1$, with confidence levels of $\approx 87\%$ (high background) and $\approx 96\%$ (low background).

\textbf{\emph{In conclusion, we find evidence of a density discontinuity at the inner side of the SE relic, with a confidence $\approx 87\%$. The density discontinuity has a best-fit compression factor of $1.62_{-0.27}^{+0.32}$, corresponding to a putative shock of Mach number $1.43_{-0.20}^{+0.23}$.}}

\subsubsection{Aperture of the SE density discontinuity}

\begin{figure*}
 	\centering
		\includegraphics[width=0.47\textwidth,keepaspectratio=true,clip=true]{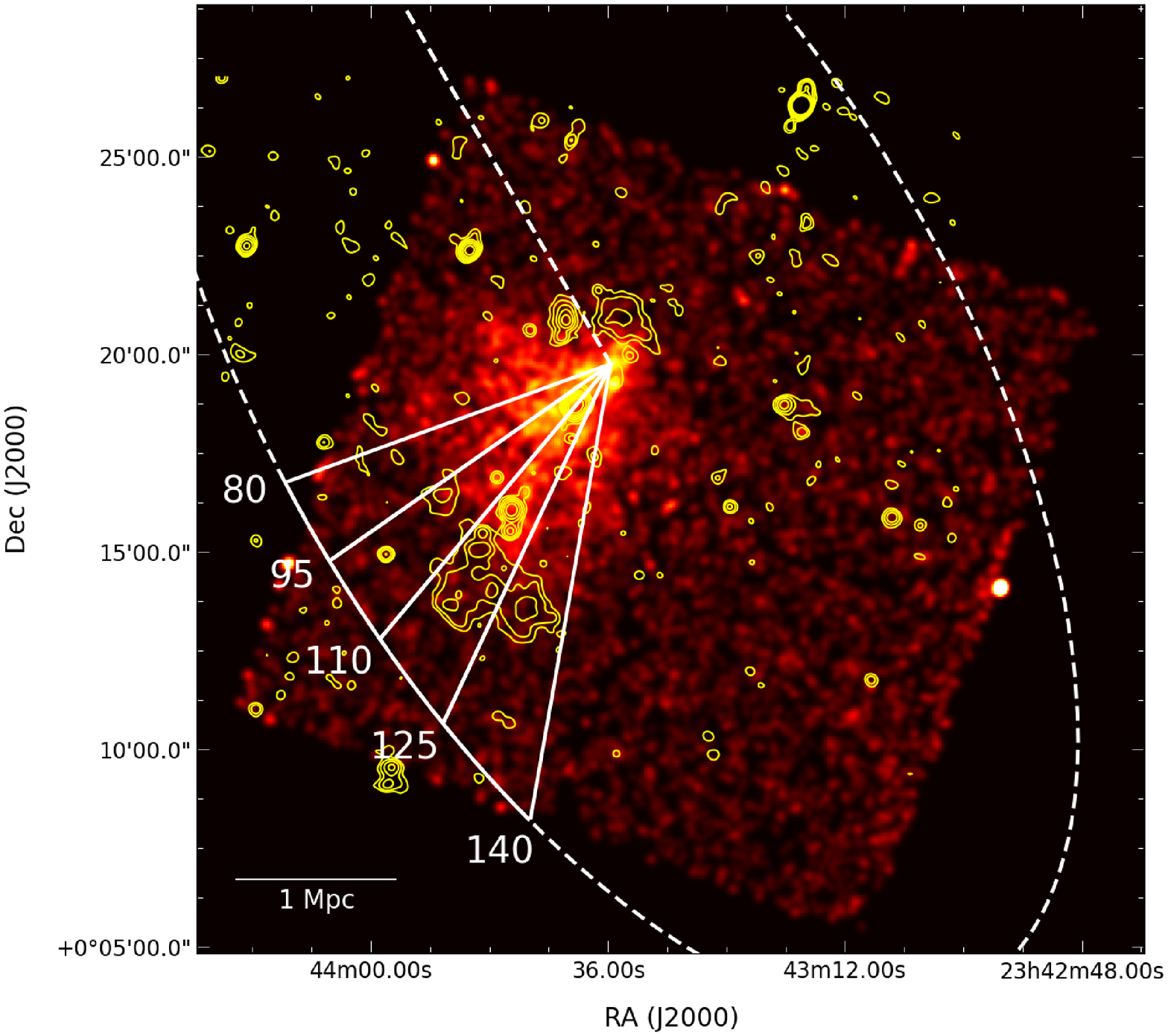} \hspace{0.05\textwidth}
  		\includegraphics[width=0.47\textwidth,keepaspectratio=true,clip=true]{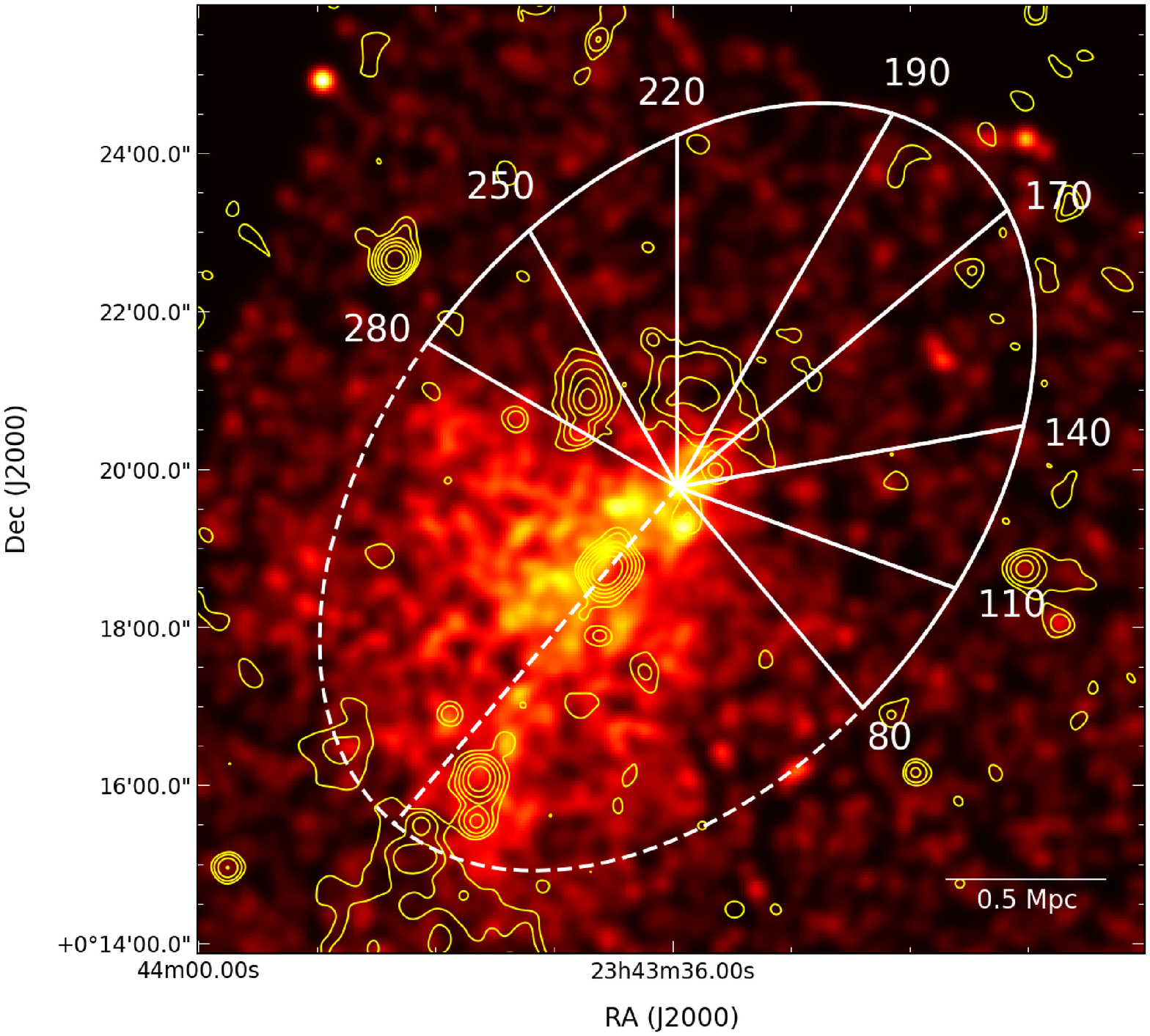}
	\caption{\chandra\ surface brightness map in the energy band 0.5-3~keV, same as in Figure \ref{fig:prettypictures}. Overlaid are the narrow elliptical wedges used to determine the extent of the SE (left) and NW (right) density discontinuities. The position angles of the wedges are marked on the image. The straight dashed lines show the positive parts of the major axes of the ellipses.}
	 \label{fig:ns-sectors}
\end{figure*}

\begin{figure}
 	\begin{center}
  		\includegraphics[width=1.1\columnwidth,keepaspectratio=true,clip=true]{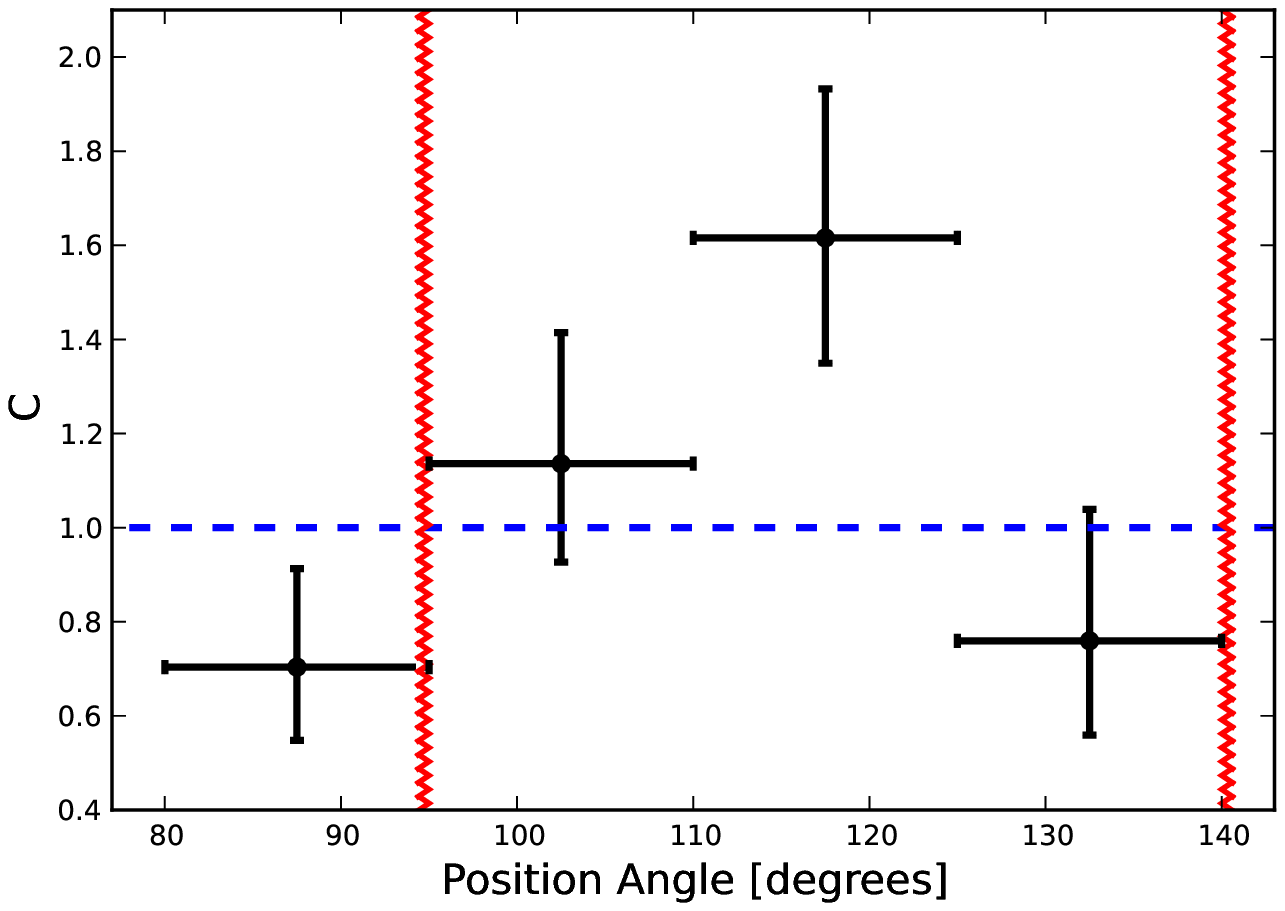}
 	\end{center}
	\caption{Density compression across the SE relic, as a function of position angle. Error bars show the angular opening of the sectors in which the density compression was calculated, and the $1\sigma$ statistical errors on the density compression. The blue line corresponds to $C=1$, and therefore data points with $C$ significantly above this line indicate regions in which a density discontinuity is present near the relic. The position angles corresponding to the tips of the relic are marked with red zig-zag lines.}
	 \label{fig:s-alpha-beta}
\end{figure}

The SE elliptical sector in Figure \ref{fig:sectors} spans position angles (PAs) between $110$ and $125$ degrees. The angles are measured from the major axis of the ellipse, which is rotated by $120$ degrees with respect to the right ascension. In the region $110 \le {\rm PA} \le 125$, the surface brightness discontinuity can be seen by eye. However, the relic's extent is in fact almost three times larger, between PAs of $95$ and $140$ degrees. To examine whether the density discontinuity spans the full relic, we extracted surface brightness profiles across the full SE relic in narrow elliptical wedges with opening angles of $15$ degrees (shown in Figure \ref{fig:ns-sectors}), binned them to a minimum of 5 counts/bin, and fitted them with a broken power-law model. In the fits, the radius of the density discontinuity and the sky background level were fixed to the best-fit values determined from the sector $110 \le {\rm PA} \le 125$ (see the previous subsection). The \xmm\ and \chandra\ profiles were fitted in parallel, using Cash statistics. 

In Figure \ref{fig:s-alpha-beta} we show the best-fit density compression as a function of PA. In the sector $110 \le {\rm PA} \le 125$, a discontinuity is detected at a confidence level of $\approx 92\%$. However, there is no immediate evidence of the shock front extending across the whole length of the radio relic; at PAs in the ranges $95-110$ and $125-140$, the density compression is consistent with 1. For the best-fit sky background values, a density compression of $1.62$ at $r_{\rm d}=10.59$ arcmin is excluded at confidence levels of $88\%$ for $95 \le {\rm PA} \le 110$ and at $98\%$ for $125 \le {\rm PA} \le 140$. Varying the discontinuity radius within its best-fit $1\sigma$ limits has only a small effect on the quoted confidence levels. To evaluate the effect of systematic uncertainties introduced by the fixing the sky background levels, we repeated the fits in the regions $95\le {\rm PA} \le 110$ and $125 \le {\rm PA} \le 140$ after simultaneously lowering (and then raising) the \xmm\ and \chandra\ sky background levels to their $-1\sigma$ $(+1\sigma)$ boundaries. For the high background levels, a density compression equal to the one determined in the region $110 \le {\rm PA} \le 125$ for the same background levels, $C=1.57_{-0.29}^{+0.33}$, is excluded at a confidence level of $90\%$ for $95 \le {\rm PA} \le 110$ and at $99\%$ for $125 \le {\rm PA} \le 140$. For the low background levels, a density compression of $C=1.65_{-0.25}^{+0.31}$ -- equal to the one determined in the region $110 \le {\rm PA} \le 125$ for the same background levels -- is excluded with a confidence of $85\%$ for $95 \le {\rm PA} \le 110$ and at $94\%$ for $125 \le {\rm PA} \le 140$.

\emph{\textbf{Therefore, the density discontinuity discovered at the SE relic appears to trace only the central part of the relic. We find moderate evidence ($\approx 85\%$) that the isolated NE part of the relic does not trace a density discontinuity of the same magnitude as the discontinuity detected across the central part of the relic. We find stronger evidence ($\approx 94\%$) that also the SW part of the relic does not trace a discontinuity of the same magnitude.}}

\begin{figure*}
 	\begin{center}
  		\includegraphics[width=0.495\textwidth,keepaspectratio=true,clip=true]{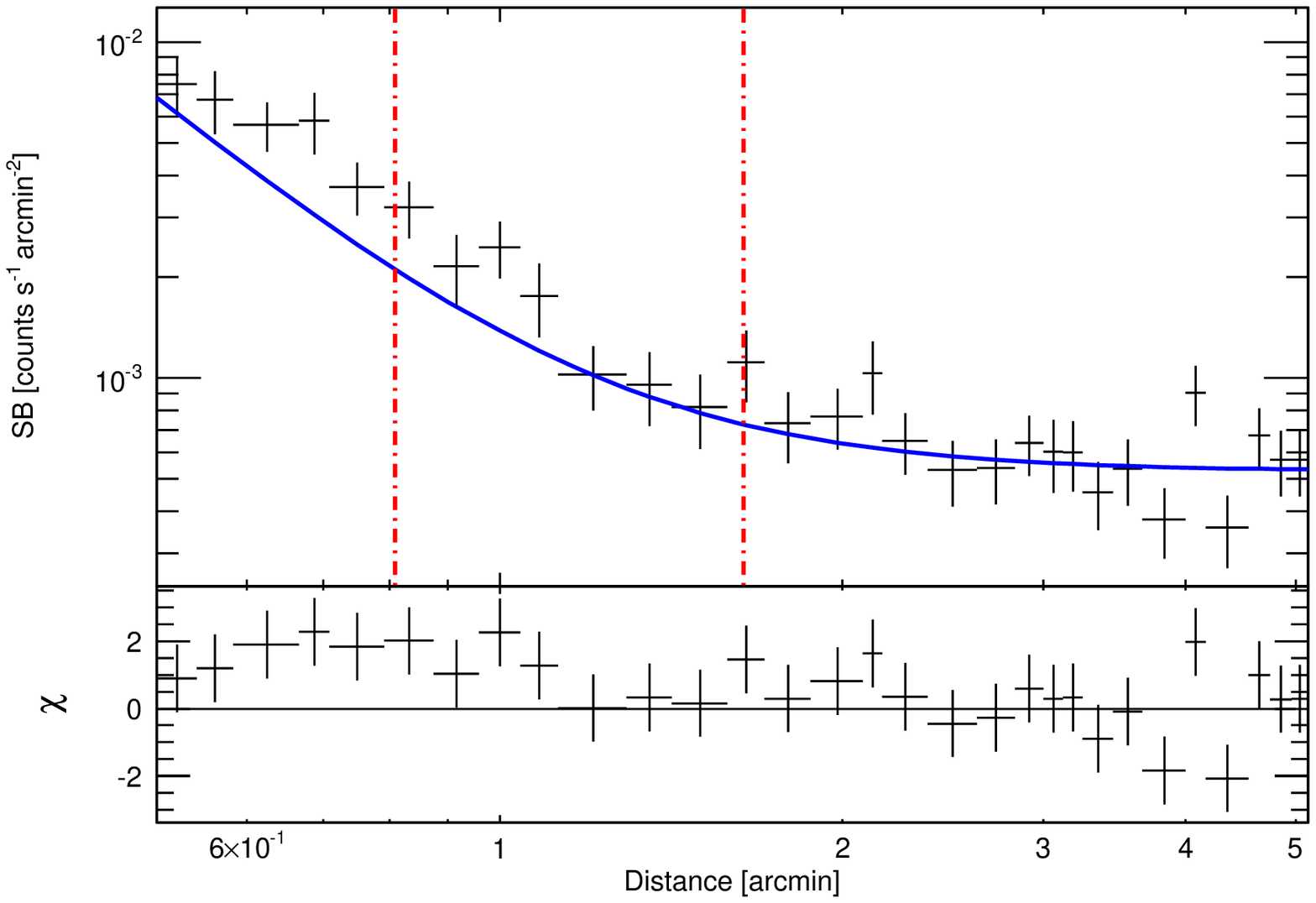}
		\includegraphics[width=0.495\textwidth,keepaspectratio=true,clip=true]{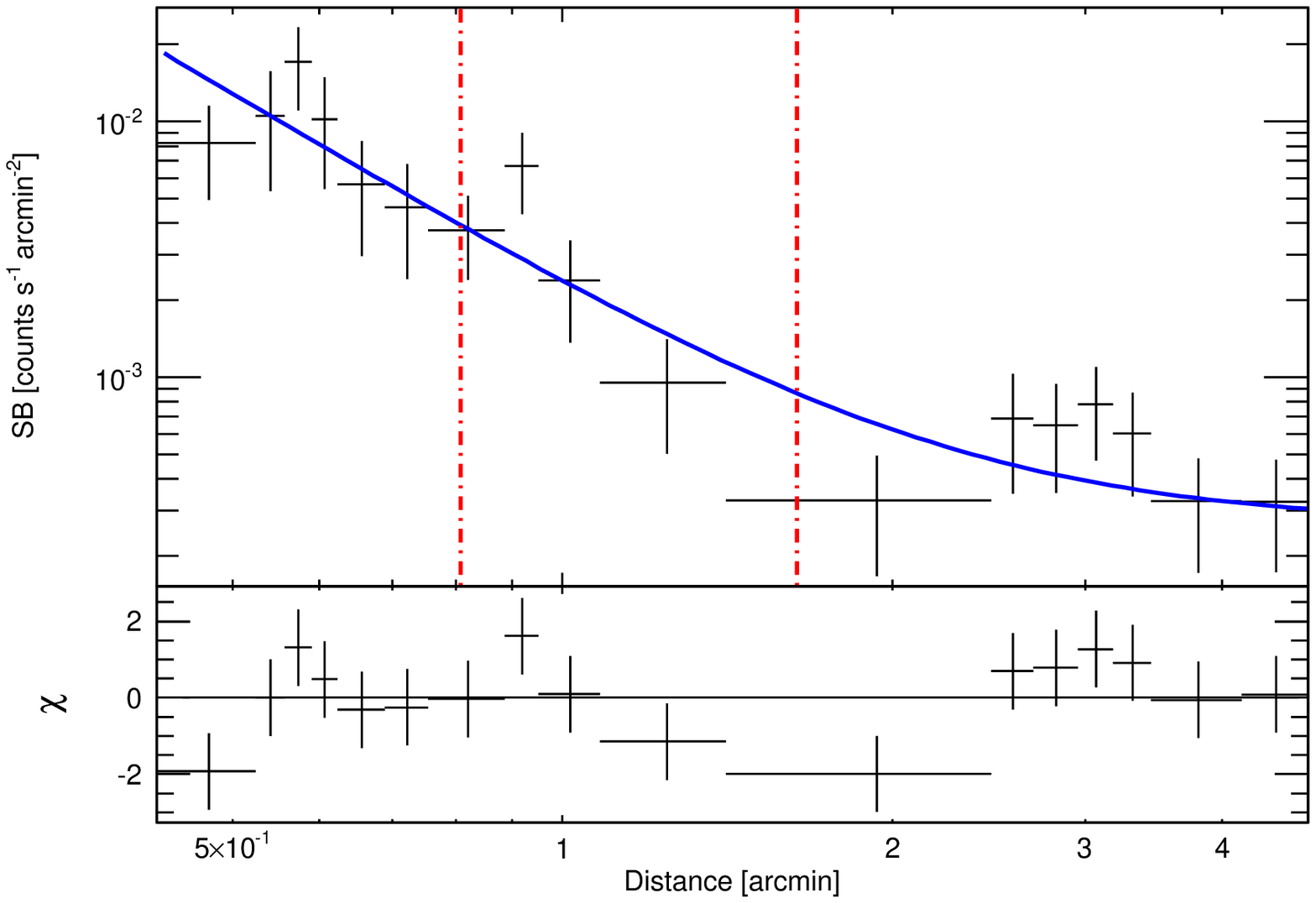}
		\includegraphics[width=0.495\textwidth,keepaspectratio=true,clip=true]{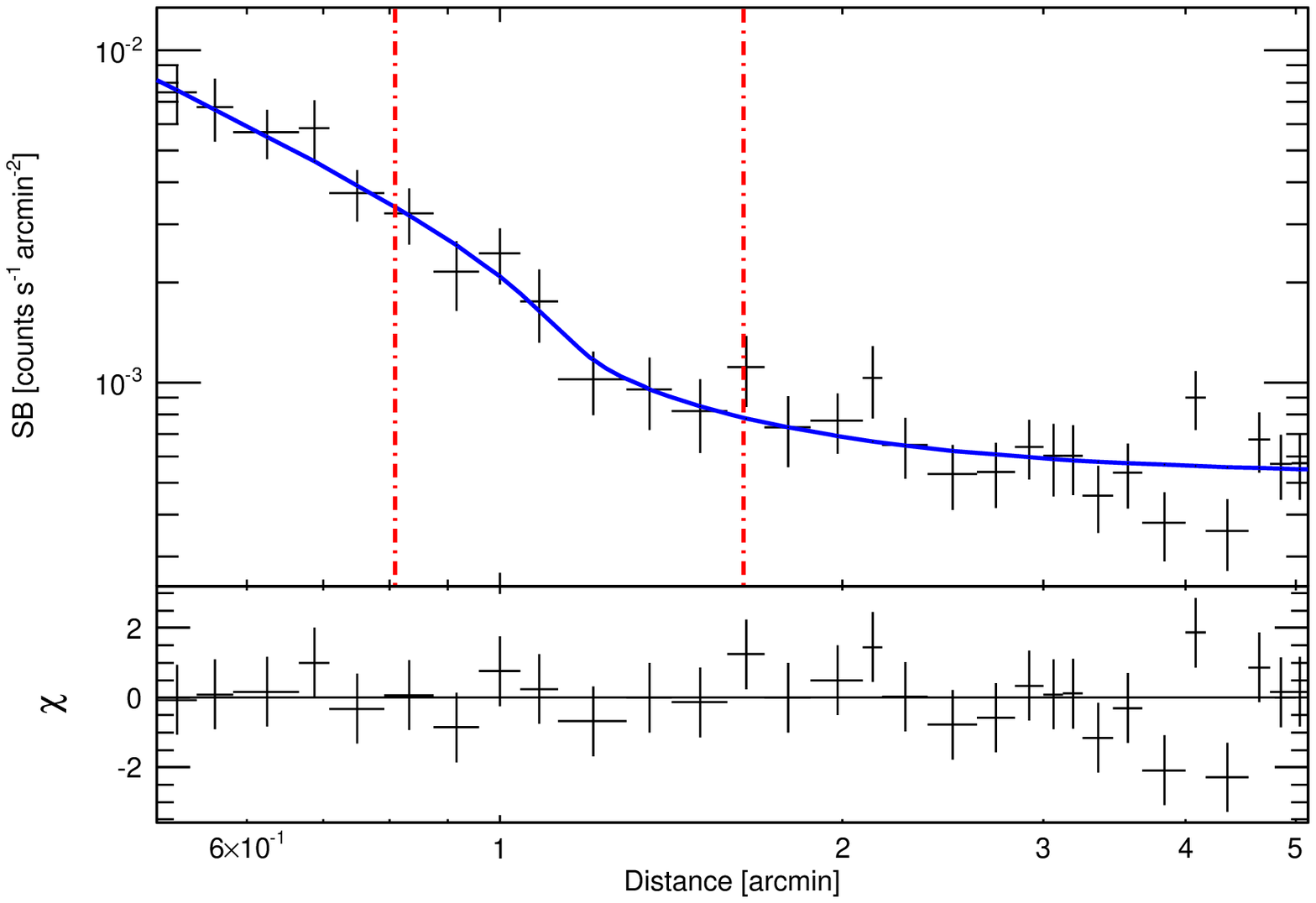}
		\includegraphics[width=0.495\textwidth,keepaspectratio=true,clip=true]{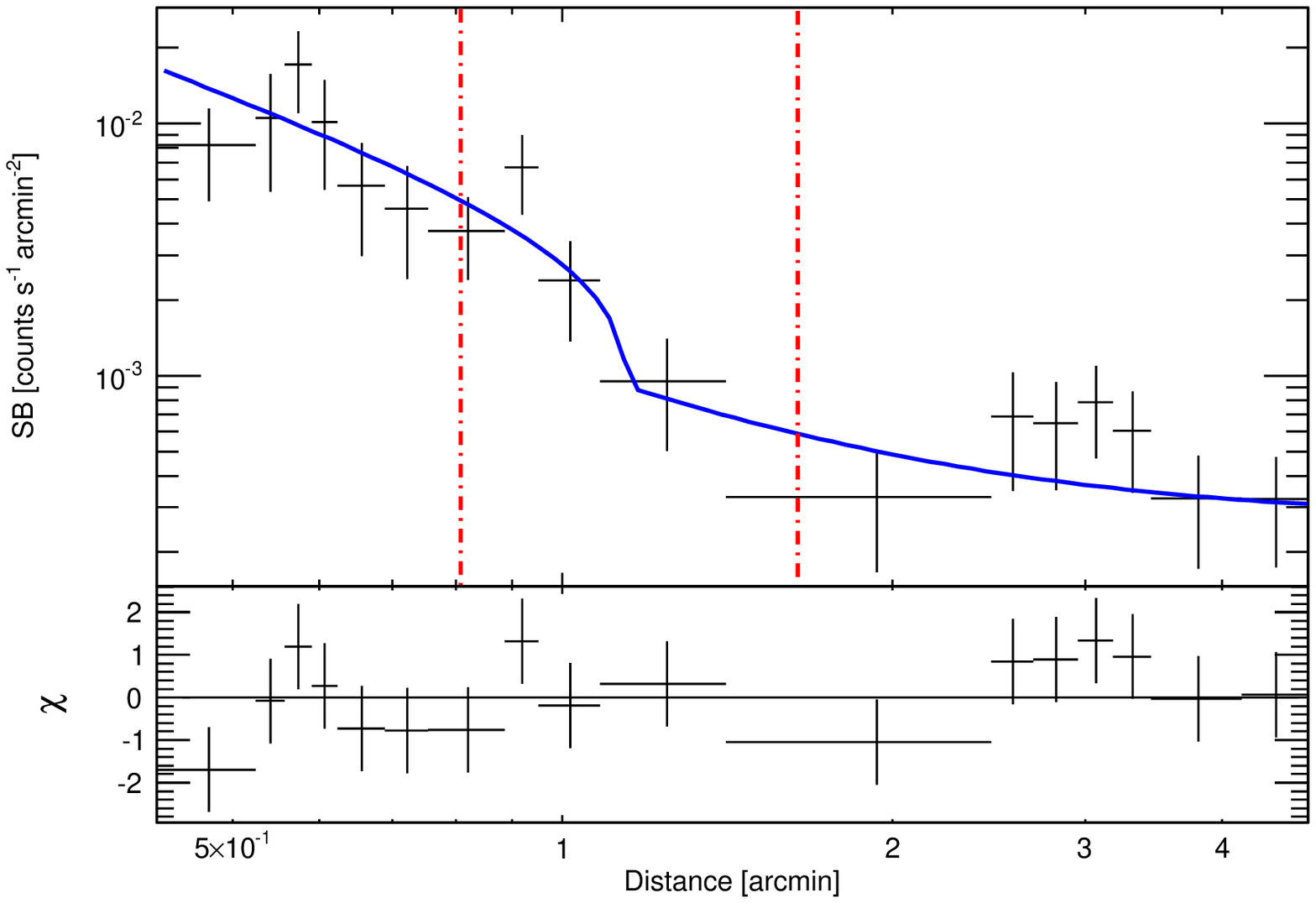}
 	\end{center}
	\caption{Best-fit models, overplotted onto the \xmm\ and \chandra\ surface brightness profiles across the NW relic. The top panels show the fits with a simple power-law model, while the bottom panels show the fits with a broken power-law model. The parameters for each of these models are presented in Table \ref{tab:n-best-fits}. Red lines correspond to the different edges of the relics (see Figure \ref{fig:sectors}), and are drawn at distances of $0.81$ and $1.64$ arcmin. \xmm\ profiles are in the left panels, and \chandra\ in the right panels. For visual clarity, the \chandra\ profiles were regrouped to a coarser binning.}
	 \label{fig:n-best-fits}
\end{figure*}

\begin{table*}
\caption{Best-fit parameters of the models fitted in parallel to the \xmm\ and \chandra\ surface brightness profiles across the NW relic. The fits were done using Cash statistics, hence we do not provide a goodness of the fit.}
\label{tab:n-best-fits}
\centering
\begin{threeparttable}
\begin{tabular}{cccccc}
 \multicolumn{6}{c}{{\sc NW Sector}} \\
 \multicolumn{6}{c}{{\sc Single Power-Law Model}} \\
 \hline
     	\multicolumn{4}{c}{$\gamma$} & ${\rm SB}_{0,\,\,{\rm XMM}}$ (cts s$^{-1}$ arcmin$^{-2}$) & ${\rm SB}_{0,\,\,{\rm Chandra}}$  (cts s$^{-1}$ arcmin$^{-2}$) \\ 
 \hline
	\multicolumn{4}{c}{$2.57_{-0.21}^{+0.54}$} & $2.97_{-2.55}^{+2.36} \times 10^{-4}$ & $(2.12\pm 0.32) \times 10^{-3}$ \\
\hline
    & & & & \\
 \multicolumn{6}{c}{{\sc Broken Power-Law Model}} \\
 \hline
     	$\alpha_1$ & $\alpha_2$ & $r_{\rm d}$ (arcmin) & $C$ &  ${\rm SB}_{0,\,\,{\rm XMM}}$ (cts s$^{-1}$ arcmin$^{-2}$) & ${\rm SB}_{0,\,\,{\rm Chandra}}$  (cts s$^{-1}$ arcmin$^{-2}$) \\ 
 \hline
	$1.28_{-0.29}^{+0.24}$ & $1.45_{-0.37}^{+0.43}$ & $1.14_{-0.09}^{+0.08}$ & $2.35_{-0.71}^{+1.30}$ & $1.64_{-0.76}^{+1.17} \times 10^{-3}$ & $3.01_{-1.03}^{+1.64} \times 10^{-3}$ \\
\hline
\end{tabular}
	\begin{tablenotes}
		\item[$\dagger$] fixed parameter
	\end{tablenotes}
   \end{threeparttable}
\end{table*}

\subsubsection{NW density discontinuity}

The surface brightness profiles across the NW relic were created in an elliptical sector with an opening angle of 65 degrees (shown in Fig. \ref{fig:sectors}). While the sector covers only the W part of the relic, a density discontinuity is not clear, by eye, across the relic's E end. 

The sky background level was determined by binning the profiles to a minimum of 5 counts/bin, and fitting a constant to the outer data bins of each profile in the radius range $6-8$ arcmin; in this radius range, the emission is essentially flat. As for the SE profile, the model fitted to the background is the sum of a constant and the instrumental background surface brightness profile. The fit was done using Cash statistics. For the \xmm\ and \chandra\ sky background levels, we found $5.27_{-0.33}^{+0.34} \times 10^{-4}$ and $2.68_{-1.06}^{+1.13}\times 10^{-4}$ cts~s$^{-1}$~arcmin$^{-2}$, respectively.

We attempted to fit the \xmm\ and \chandra\ surface brightness profiles with a simple power-law and with a broken power-law, in the radius range $0.5-5$ arcmin. The \xmm\ and \chandra\ profiles were binned to a minimum of 5 counts/bin. The fits were done using Cash statistic. The best-fit parameters for the two models are very similar regardless of the binning, and the same is true for the difference in Cash statistic between the two models. We assumed that the same physical model describes both profiles, and fitted in parallel the data from the two satellites; therefore, all the parameters describing cluster emission, with the exception of the central surface brightness values, were linked in the fits.

\begin{figure*}
	\centering
		\includegraphics[width=0.495\textwidth, keepaspectratio=true, clip=true]{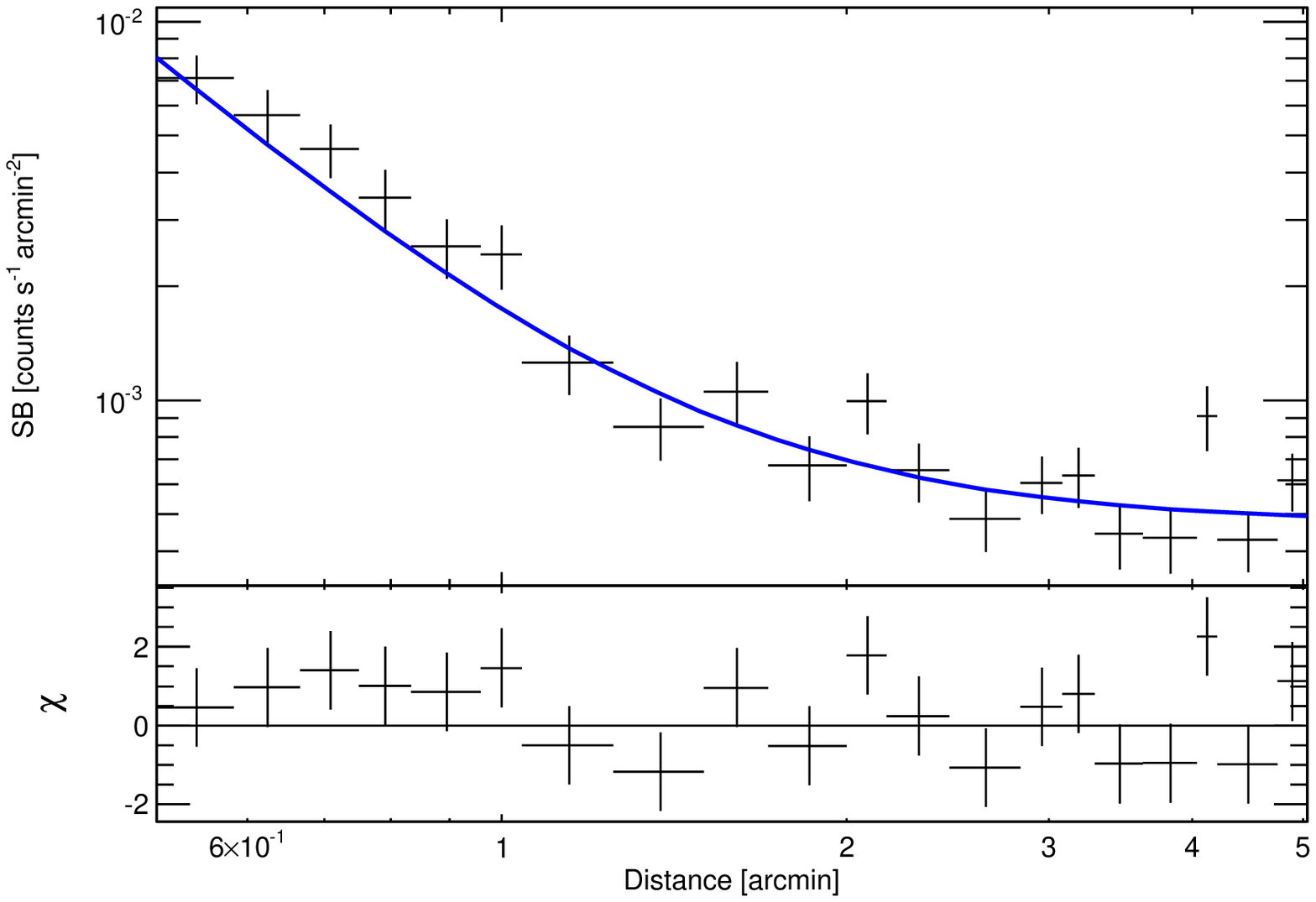}
		\includegraphics[width=0.495\textwidth, keepaspectratio=true, clip=true]{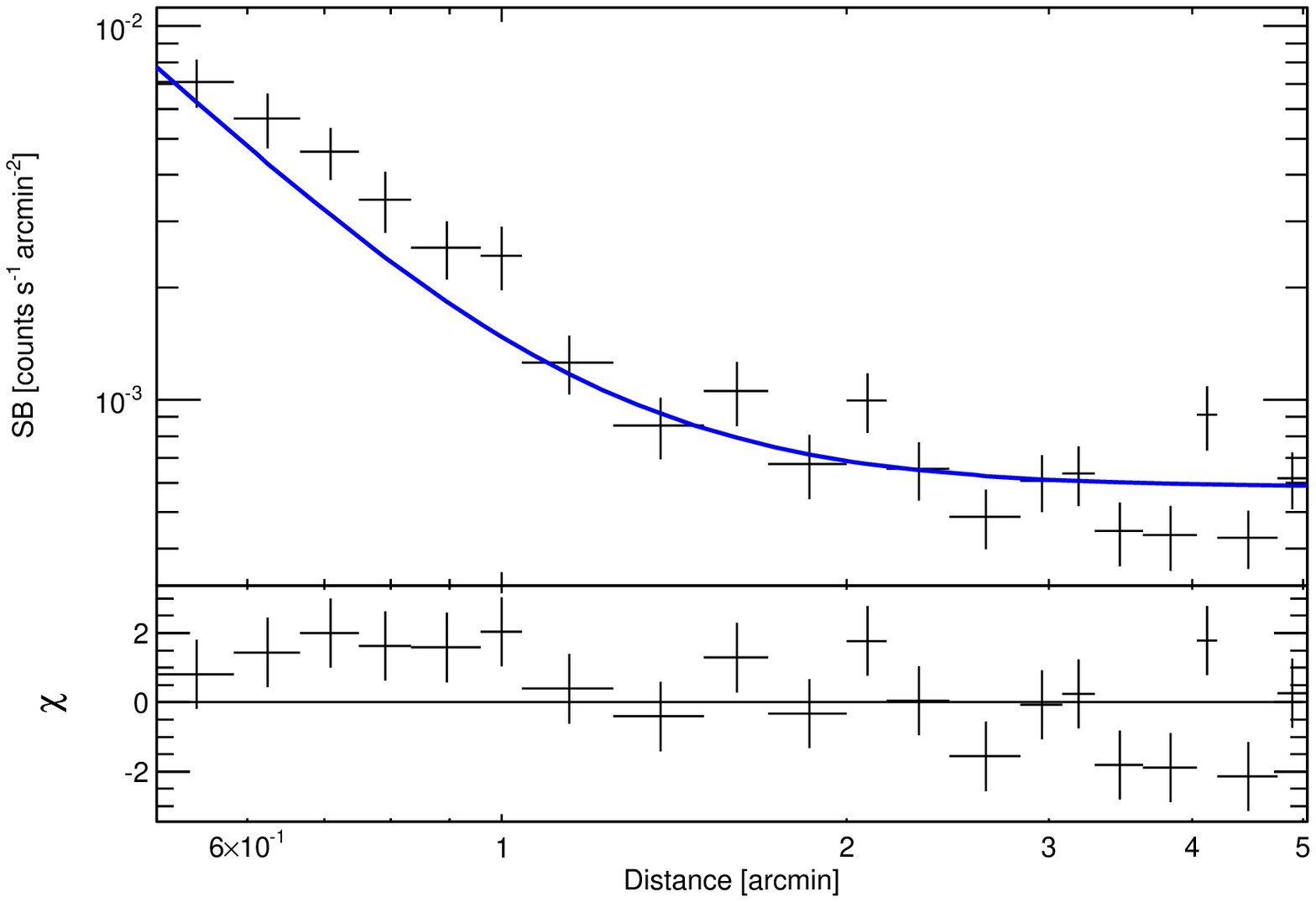}
		\includegraphics[width=0.495\textwidth, keepaspectratio=true, clip=true]{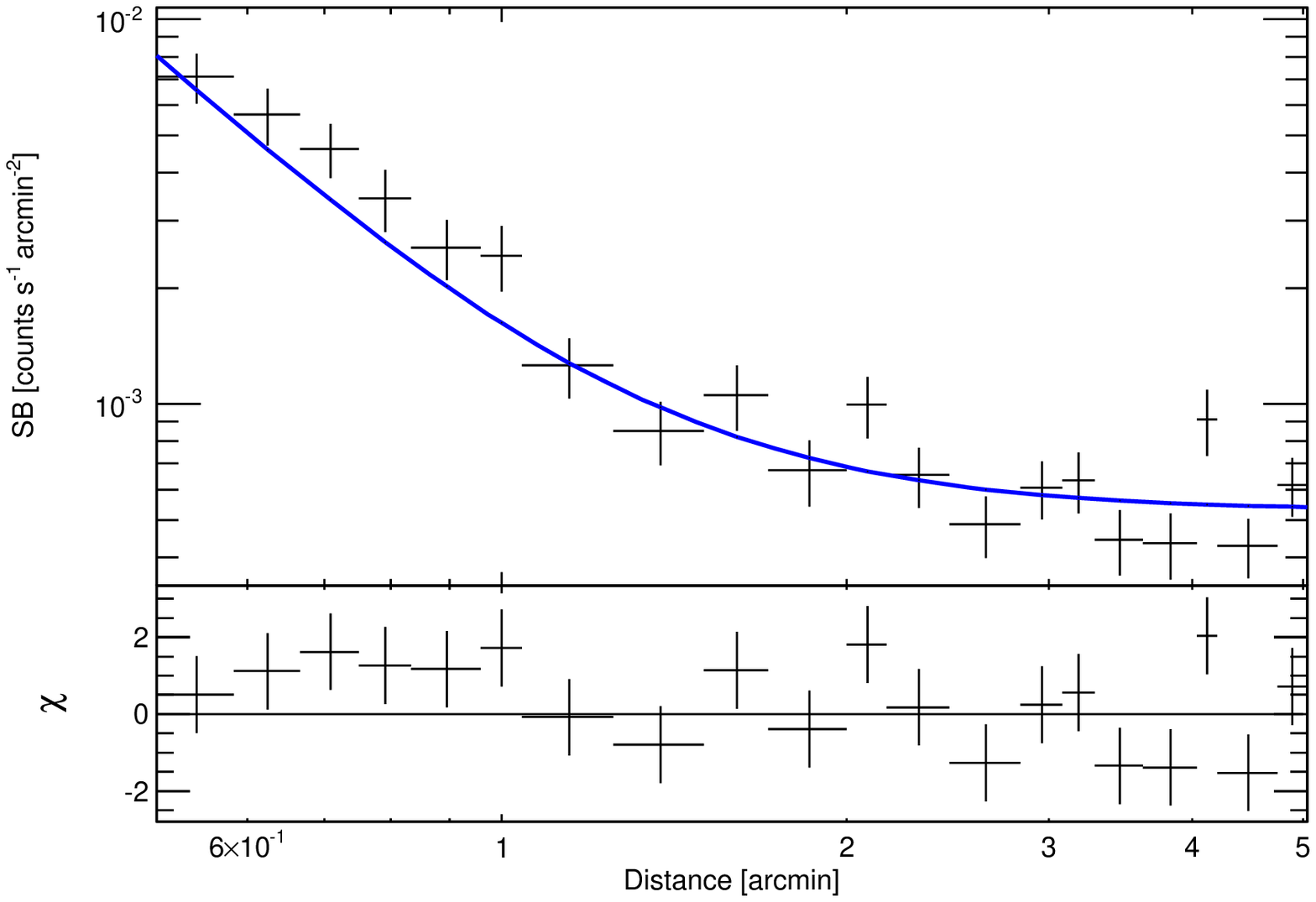} \hspace{0.04\textwidth}
		\begin{minipage}[t][-0.5cm][b]{0.45\textwidth}
			\captionsetup{justification=raggedright}
			\caption{Best-fit power-law models fitted exclusively to the \xmm\ NW surface brightness profile. At the bottom is the model with a fixed sky background level of $5.27\times 10^{-4}$~cts~s$^{-1}$~cm$^{-2}$, which is the best-fit background level determined from the outer bins of the profile. The top-left and top-right panels show the best-fit power-law model for background levels of, respectively, $4.73\times 10^{-4}$ and $5.84\times 10^{-4}$~cts~s$^{-1}$~arcmin$^{-2}$ -- the boundaries of the $90\%$ confidence interval for the \xmm\ sky background level. The sky background levels were fixed in the fits. The fits were done in the radius range $0.5-5.0$ arcmin.}
			\label{fig:xmm-n-pow}
		\end{minipage}
\end{figure*}

\begin{table*}
\caption{Best-fit parameters of the broken power-law models refitted in parallel to the surface brightness profiles across the NW relic after simultaneously varying the \xmm\ and \chandra\ sky background levels by $1\sigma$. The fits were done using Cash statistics, hence we do not provide a goodness of the fit.}
\label{tab:n-robustness}
\centering
\begin{tabular}{cccccc}
 \multicolumn{6}{c}{{\sc NW Sector}} \\
 \multicolumn{6}{c}{{\sc Low Background Level -- Broken Power-Law}} \\
 \hline
     	$\alpha_1$ & $\alpha_2$ & $r_{\rm d}$ (arcmin) & $C$ &  ${\rm SB}_{0,\,\,{\rm XMM}}$ (cts s$^{-1}$ arcmin$^{-2}$) & ${\rm SB}_{0,\,\,{\rm Chandra}}$  (cts s$^{-1}$ arcmin$^{-2}$) \\ 
 \hline
 	$1.27_{-0.30}^{+0.24}$ & $1.92_{-0.59}^{+0.91}$ & $1.14_{-0.10}^{+0.09}$ & $2.18_{-0.83}^{+1.71}$ & $1.68_{-0.79}^{+1.25} \times 10^{-3}$ & $2.98_{-1.03}^{+1.70} \times 10^{-3}$ \\
\hline
    & & & & \\
 \multicolumn{6}{c}{{\sc High Background Level -- Broken Power-Law}} \\
 \hline
     	$\alpha_1$ & $\alpha_2$ & $r_{\rm d}$ (arcmin) & $C$ &  ${\rm SB}_{0,\,\,{\rm XMM}}$ (cts s$^{-1}$ arcmin$^{-2}$) & ${\rm SB}_{0,\,\,{\rm Chandra}}$  (cts s$^{-1}$ arcmin$^{-2}$) \\ 
 \hline
	$1.33_{-0.24}^{+0.29}$ & $1.17_{-0.26}^{+0.27}$ & $1.16\pm 0.08$ & $2.50_{-0.70}^{+1.20}$ & $1.42_{-0.70}^{+1.03} \times 10^{-3}$ & $2.87_{-1.01}^{+1.52} \times 10^{-3}$ \\
\hline
\end{tabular}
\end{table*}

For the simple power-law fit, the power-law indices were linked for the \xmm\ and \chandra\ models, but were free in the fit. The sky background levels were kept fixed to the values determined from the outer bins of the corresponding sectors. The best-fit parameters are summarized in Table \ref{tab:n-best-fits}, and the model is shown in Figure \ref{fig:n-best-fits}. None of the profiles is very well described by a power-law. For the \chandra\ profile, the model overestimates the surface brightness between $\sim 1.0-2.5$ arcmin, while for the \xmm\ profile, the model underestimates the surface brightness at radii $\lesssim 1$ arcmin. It can be seen by eye that the \chandra\ surface brightness profile is not truly a power-law. For the \xmm\ profile, the underestimated surface brightness at small radii could, presumably, be an effect of fitting the data from the two satellites in parallel. Therefore, we fitted the \xmm\ profile separately, taking again into account the satellite's PSF and fixing the sky background level to $5.27\times 10^{-4}$~cts~s$^{-1}$~arcmin$^{-2}$. The parameters of the power-law that best describes the \xmm\ profile are $\gamma = 2.44_{-0.16}^{+0.26}$ and ${\rm SB}_{0, \,\,{\rm XMM}} = 5.00_{-2.50}^{+1.98}\times 10^{-4}$~cts~s$^{-1}$~arcmin$^{-2}$, but the new model, presented in Figure \ref{fig:xmm-n-pow}, does not provide a good description of the \xmm\ profile either. We also varied the \xmm\ sky background level to its $\pm 1\sigma$ and then $90\%$-confidence boundaries, and tried refitting the profile with the new background levels. Yet, this also did not considerably improve the fit. Hence, we conclude that none of the surface brightness profiles is satisfactorily fitted by a broken power-law model.

The single power-law fits suggest that the data might be fitted better by a broken power-law model. Therefore, we considered a broken power-law model with density compression $1\le C < 4$. As before, the \xmm\ and \chandra\ profiles were fitted in parallel in the radius range $0.5-5.0$ arcmin, and the sky background levels were fixed to the corresponding best-fit values. The best-fit parameters are listed in Table \ref{tab:n-best-fits}, and the models are shown in Figure \ref{fig:n-best-fits}. The best-fit compression factor is $C=2.35_{-0.71}^{+1.30}$, and it is greater than 1 at a confidence level of $96\%$. If the discontinuity is a shock, the compression factor indicates a Mach number of $2.06_{-0.76}^{+1.39}$. 

We compared the best-fit simple power-law and broken power-law models using the likelihood-ratio test. The difference in Cash statistic between the two models is $\Delta C = 8.45$, with $3$ degrees of freedom (d.o.f.); the cumulative distribution function of the chi-squared distribution is $\mathcal{F}_{\rm CDF} (\Delta C, n_{\rm d.o.f.}) = 96.24\%$ for $\Delta C=8.45$ and $n_{\rm d.o.f.} = 3$. Therefore, the broken power-law model provides a better description of the data compared to the simple power-law model with a confidence of $\approx 96\%$.

We again verified the robustness of our results by simultaneously varying the \chandra\ and \xmm\ sky background levels to their $\pm 1\sigma$ limits, and refitting in parallel the surface brightness profiles with the broken power-law density model. The fits are summarized in Table \ref{tab:n-robustness}. For both sets of sky background levels, the density compression remains $>1$ with a confidence of $\approx 82\%$ for the high background levels, and of $\approx 99\%$ for the low background levels.

\emph{\textbf{In conclusion, we find moderate evidence (confidence level $\approx 82\%$) for a density discontinuity at the NW relic. The density discontinuity has a best-fit compression factor of $2.35_{-0.71}^{+1.30}$, corresponding to a putative shock of Mach number $2.06_{-0.76}^{+1.39}$.}}

\subsubsection{Aperture of the NW density discontinuity}

\begin{figure}
 	\begin{center}
  		\includegraphics[width=1.1\columnwidth,keepaspectratio=true,clip=true]{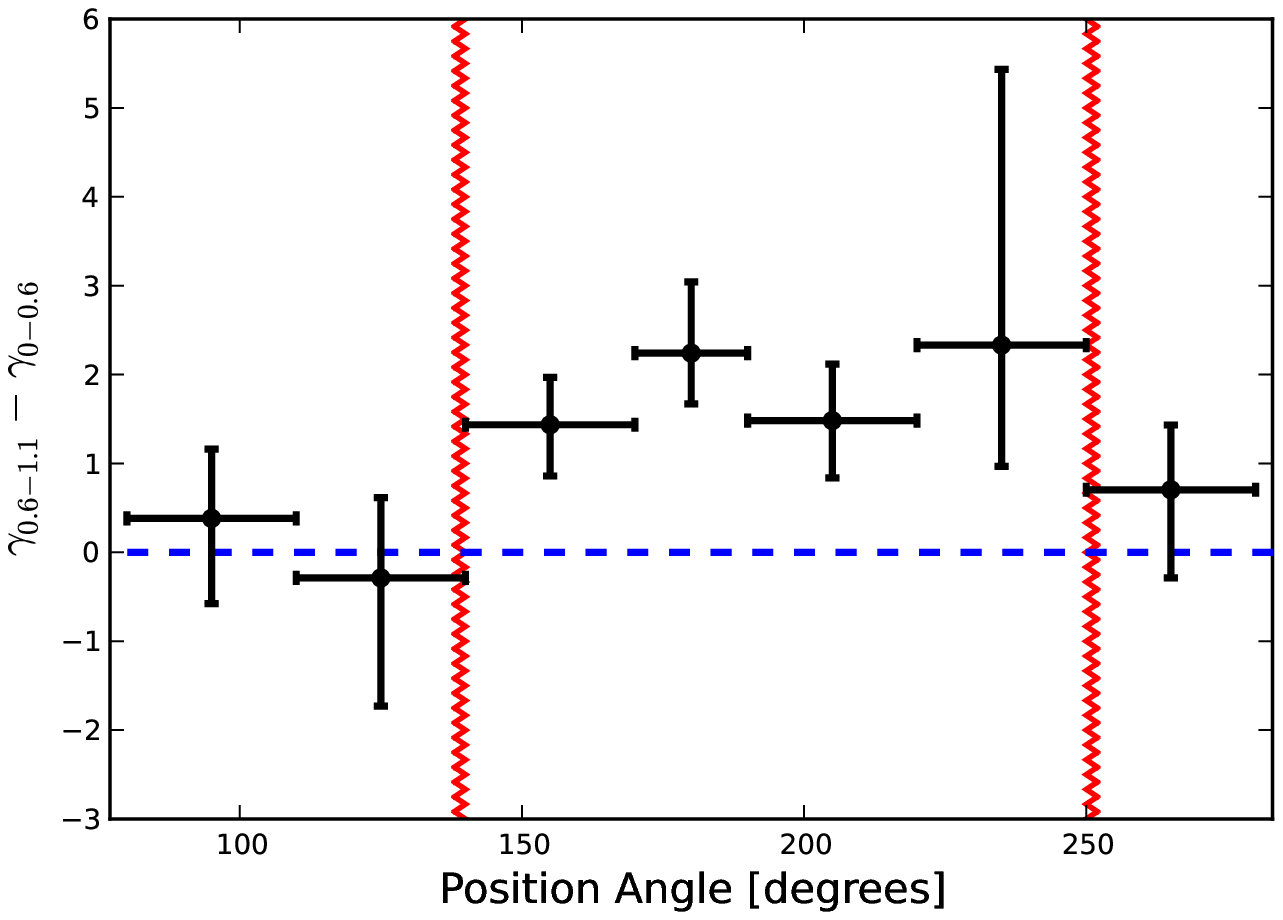}
 	\end{center}
	\caption{Difference in slope between two power-laws fitted to surface brightness profiles extracted from narrow wedges across the NW relic, plotted as a function of position angle. The power-laws were fitted in the radii ranges $0-0.6$ and $0.6-1.1$ arcmin. Error bars show the angular opening of the wedges, and the propagated $1\sigma$ statistical errors on the slope difference. The blue line corresponds to equal slopes, and therefore data points with $\gamma_{0.6-1.1}\, -\, \gamma_{0-0.6}$ significantly above this line indicate regions in which a kink is present near the relic. The position angles corresponding to the tips of the relic are marked with red zig-zag lines.}
	 \label{fig:n-alpha-beta}
\end{figure}

The NW elliptical sector in Figure \ref{fig:sectors} spans PAs between $145$ and $210$ degrees, measured from the major axis of the ellipse (the major axis of the ellipse is rotated by 230 degrees from the right ascension). To examine more carefully the extent of the density discontinuity detected at the NW relic, we extracted surface brightness profiles in narrow elliptical wedges between PAs of $80$ and $280$ degrees (see Figure \ref{fig:ns-sectors}), and fitted them with power-laws in the radii ranges $0-0.6$ and $0.6-1.1$ arcmin. As can be seen in Figure \ref{fig:n-best-fits}, power-laws fitted in these radii ranges to the \xmm\ and \chandra\ profiles are expected to have different slopes in wedges in which a discontinuity is present near the best-fit jump distance of $r_{\rm d} = 1.14_{-0.09}^{+0.08}$ determined at the NW relic. In all the fits, the sky background level was fixed to the best-fit values determined in the sector with $145\le PA \le 210$ (see the previous subsection). The fits were done using Cash statistics. In Figure \ref{fig:n-alpha-beta} we plot the slope difference between the radii ranges $0.6-1.1$ and $0-0.6$ arcmin, as a function of PA. There is a significant difference in slope between PAs of $140$ and $250$ degrees, while at smaller and larger angles, the difference is consistent with 0; PAs of $140$ and $250$ degrees correspond to the tips of the NW relic.

\textbf{\emph{Therefore, we conclude that the NW relic, unlike its SE counterpart, traces a density discontinuity along its full length.}}

\section{Discussion}
\label{sec:discussion}

\subsection{The merger in ZwCl~2341.1+0000}

The \xmm\ and \chandra\ images reveal a NW-SE elongated merging cluster, with the peak X-ray emission located near the smaller, NW relic. The position of the radio relics with respect to the apparent merger axis (the NW-SE line following the brightest cluster emission) suggest an off-axis cluster merger; based on the numerical simulations of double-relic clusters presented by \citet{vanWeeren2011b}, we estimate an impact parameter $\gtrsim 4-5$ times the core radius of the more massive cluster. In terms of the X-ray and radio morphologies, the cluster bears some resemblance with another Zwicky cluster -- ZwCl~0008.8+5215 \citep{vanWeeren2011c}. Optically, however, ZwCl~2341.1+0000 might have a more complex structure, with three or more distinct subclumps \citep{vanWeeren2009b,Boschin2013}. This warns against using only X-ray and radio brightness maps to infer conclusions about the merger geometry. We refer the reader to the work of \citet{Boschin2013} for an extensive discussion on the merger scenario.

\subsection{The nature of the density discontinuities}

Density discontinuities in the ICM can mark either cold fronts or shocks, depending on the sign of the temperature jump across the discontinuity. Unfortunately, the archival \xmm\ and \chandra\ observations of ZwCl~2341.1+0000 are not deep enough to measure the temperatures in the outer regions of the discontinuities. There is an existing $50$ ks \emph{Suzaku} observation of the cluster, but the combination of a relatively short exposure time and of significant contamination by point sources near the discontinuities makes it impossible to accurately measure a possible temperature jump. Results from the \emph{Suzaku} observation were presented by \citet{Akamatsu2013}, who found no evidence of a temperature jump; however, their analysis did not exclude any point sources from the FOV, which likely affected the measurements. In the absence of deeper data that would allow an accurate spectral analysis, our discussion about the nature of the newly discovered density discontinuities in ZwCl~2341.1+0000 is merely qualitative. 

\begin{figure*}
 	\begin{center}
		\includegraphics[height=0.5\columnwidth,keepaspectratio=true,clip=true]{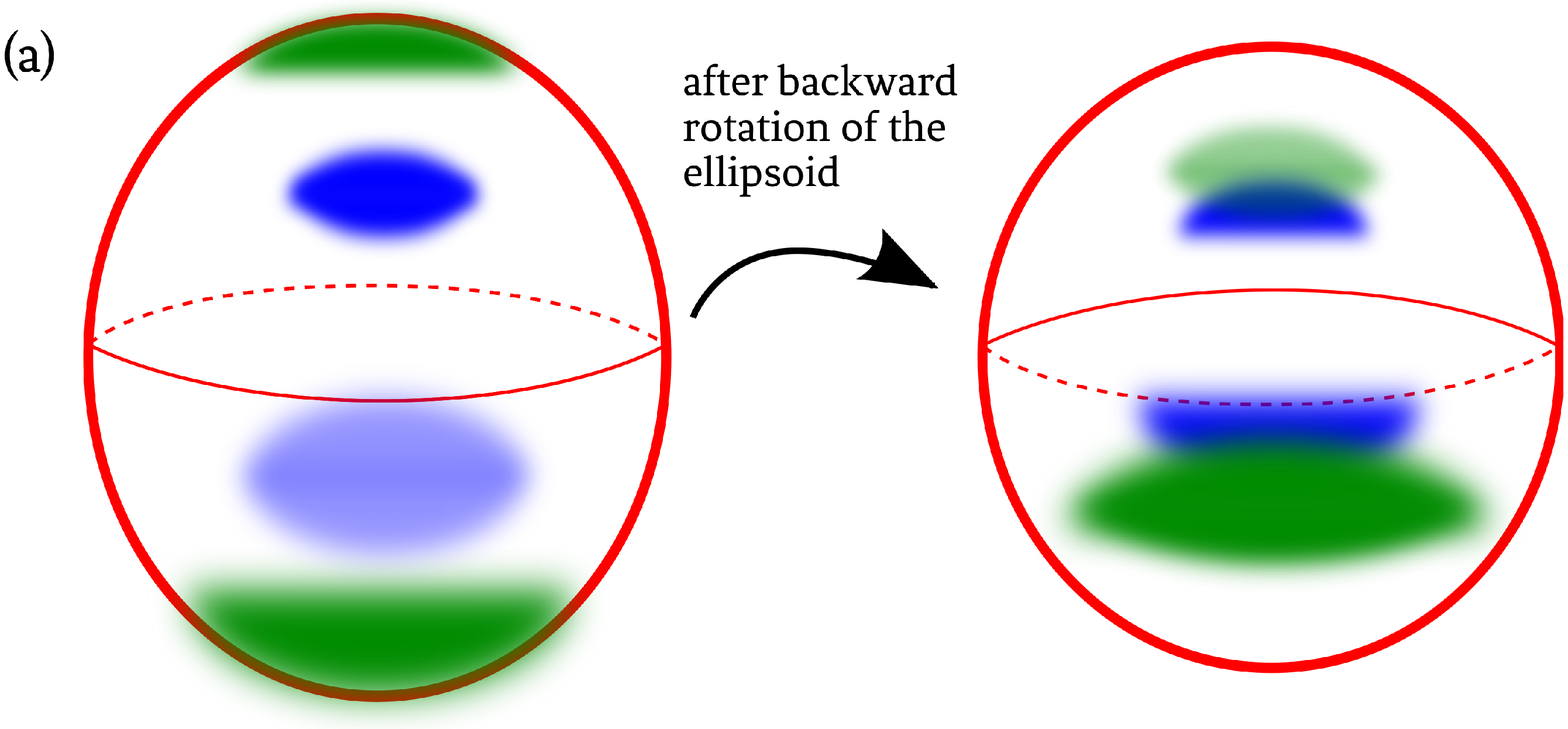}\hspace{2cm}
  		\includegraphics[height=0.5\columnwidth,keepaspectratio=true,clip=true]{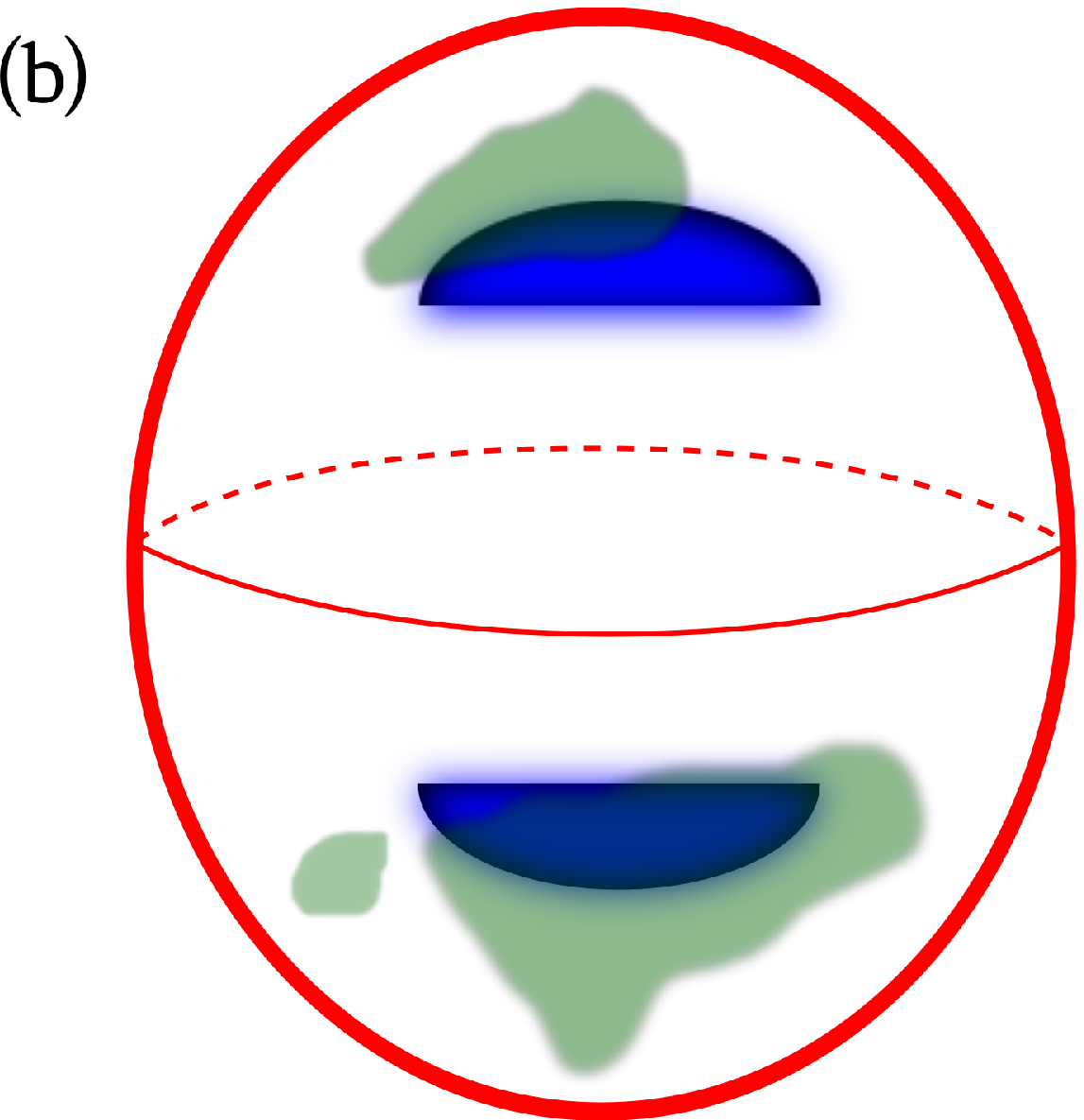}
 	\end{center}
	\caption{Drawing of possible configurations in which the radio relics would be seen, in projection, at the location of cold fronts. Cold fronts are shown in blue, and radio relics are in green. In (a), particles are accelerated by shocks at the location of the radio relics, as shown in the image on the left, which is a view of the cluster in the plane of the sky. There are interior cold fronts that are seen at some angle with respect to the plane of the sky; the S cold front is on the back side of the ellipsoidal surface, while the N cold front is on the front side. The image on the right shows the same ellipsoidal surface, at a different viewing angle (`backward' rotation refers to a rolling of the ellipsoid in a backward direction with respect to the reader). For this particular viewing angle, the radio relics are seen projected onto the cold fronts. The NW relic is seen at the back side of the ellipsoid, while the S one is on the front side. In (b), an ellipsoidal shock surface is shown in red. Particles are accelerated everywhere at the shock surface, but the particle acceleration efficiency is high enough to lead to visible radio emission only in the regions shown in green; these regions are on the shock surface. In projection, however, the relics would misleadingly appear to be associated with the cold fronts.}
	 \label{fig:geometry}
\end{figure*}

Essentially all the observational and theoretical evidence points towards radio relics being formed by particle acceleration at shock fronts \citep[e.g.][and references therein]{Brueggen2012}. Cold fronts, on the other hand, are the consequence of merger-induced sloshing in the ICM \citep[e.g.][]{Markevitch2007}. The location of the density discontinuities in ZwCl~2341.1+0000 -- coincident, in projection, with that of the relics -- suggests that the discontinuties correspond to shock fronts. Nevertheless, if the relics trace elliptical shock caps tilted with respect to the plane of the sky, then the relics could be seen, incidentally, projected onto inner cold fronts (see Fig. \ref{fig:geometry}a for a geometrical respresentation of this scenario). Alternatively, if the relics trace elliptical shock caps and the particle acceleration efficiency varies across the caps \citep{Skillman2013}, then radio emission could potentially be visible only on the side of the caps. In projection, the radio relics, seen only on the side of the shock caps, could coincidentally overlap inner density discontinuities associated with cold fronts (Fig. \ref{fig:geometry}b). In this case, there should be additional density discontinuities even further out in the cluster outskirts -- those associated with the shock fronts. We find no evidence of these discontinuities, but, even if they existed, they would be impossible to detect with the current data. While complex geometries such as those presented in Fig. \ref{fig:geometry} cannot be ruled out, no cold fronts have been observed yet so close to radio relics, and hence we consider it more likely that the discontinuities mark shock fronts.

The discussion that follows is based on the assumption that the density discontinuities correspond to shock fronts.

\subsection{Failure of the standard DSA?}

Based on the low-frequency \gmrt\ observations and on the VLA 325 and 1400 MHz observations presented by \citet{Bagchi2002}, the NW and SE radio relics have integrated spectral indices of $-0.49\pm 0.18$ and $-0.76\pm 0.17$, respectively \citep{vanWeeren2009b}. \citet{Giovannini2010} combined D-array \vla\ data with the \gmrt\ data to find integrated spectral indices of $\approx -1.2$ for both relics. For a steady-state homogeneous source, the injection spectral index is flatter than the integrated spectral index by $0.5$ \citep[e.g.][]{Blandford1987}. Indeed, this relation appears to hold well for other radio relics, such as CIZA~J2242.8+5301 \citep{vanWeeren2010,Stroe2013} and 1RXS J0603.3+4214 \citep{vanWeeren2012}. Using the results of \citet{vanWeeren2009b}, the injection spectral indices at the NW and SE radio relics would be $\approx 0.01$ and $\approx -0.26$, respectively. From the results of \citet{Giovannini2010}, the expected injection spectral indices are $\approx -0.7$ at both radio relics.

In the standard DSA model, the Mach number of the shock is related to the injection spectral index, $\alpha_{\rm inj}$, via $\mathcal{M}^2 = \frac{2\alpha_{\rm inj}-3}{2\alpha_{\rm inj}+1}$. For $\mathcal{M}$ to be a real number, $\alpha_{inj}$ is required to be $<-0.5$. This requirement fails for the spectral indices derived from the \gmrt\ data alone. For injection spectral indices of $\approx -0.7$, as measured from the \vla+\gmrt\ data, the Mach numbers of the shocks should be $\approx 3.3$. Our analysis of the archival \xmm\ and \chandra\ observations revealed density discontinuities at the NW and SE relics corresponding to Mach numbers of $2.06_{-0.76}^{+1.39}$ and $1.43_{-0.20}^{+0.23}$, respectively. Hence, the Mach number derived at the SE relic is significantly below the value predicted by the radio data.

We fitted the surface brightness profiles assuming that the ICM geometry is well-described by a prolate spheroid. If, instead, the spheroid has semi-major axes $(R_1,R_2, R_3) = (b,a,a)$, with $a \ge b$ and $R_3$ also aligned with the line of sight, then the density compression factors measured of the shock would be lower. This would only strengthen our conclusions regarding the discrepancies between the X-ray-measured and radio-derived Mach numbers. For example, for the ellipticities ($\epsilon$) of the ellipses shown in Figure \ref{fig:ns-sectors}, $\epsilon_{\rm NW} = 1.5$ and $\epsilon_{\rm SE} = 2.1$, changing the geometry from a prolate spheroid to a spheroid with semi-major axis $(R_1,R_2, R_3) = (b,a,a)$, lowers the best-fit density compression factors by $\approx 10\%$, to $C_{\rm NW} = 2.18_{-0.64}^{+1.25}$ and $C_{\rm SE} = 1.45_{-0.28}^{+0.42}$, although these values are still consistent with those derived for a prolate spheroid.

The magnitude of the density discontinuities could be underestimated if the measurements are affected by projection effects, or if the shape of the shock fronts deviates from the elliptical shapes assumed in our analysis. However, neither projection effects, nor complex shock front geometries could reconcile the X-ray-measured Mach numbers with the spectral indices derived from the \gmrt\ data. Looking at other recent observational results, reconciling the X-ray-measured Mach numbers and the Mach numbers predicted from the radio spectral indices under the assumptions of standard DSA is not always possible. In CIZA~J2242.8+5301, for example, the Mach number at the N relic is significantly lower ($\sim 2$) than predicted by the radio data ($\sim 4.5$). In another relic cluster, Abell 2256, the integrated spectral index is also very flat, $-0.81\pm 0.03$ \citep{vanWeeren2012}, suggesting an injection spectral index $\sim -0.3$. Moreover, Mach numbers $\gtrsim 4$ have not been measured in X-ray at any of the known radio relics. Therefore, in light of previous results, the discrepancy between the spectral indices and the Mach numbers in ZwCl~2341.1+0000 is not necessarily a biased result, but rather further evidence of the insufficiencies of the standard DSA model.

Nevertheless, it is also worth mentioning that X-ray-measured and radio-predicted Mach numbers are consistent in other galaxy clusters, e.g. Abell 754 \citep{Macario2011} and Abell 521 \citep{Bourdin2013}.

The flat spectral indices at the radio relics in ZwCl~2341.1+0000 could be explained if we are seeing a young merger, in which the relics have only recently lit up. In this scenario, the integrated spectral index, $\alpha$, is expected to be $>\alpha_{\rm inj}-0.5$ at low frequencies because the electrons have not yet had time to age and hence to steepen the spectrum. More steepening would be observed at higher frequencies, which could explain the steeper spectral indices reported by \citet{Giovannini2010}. The distance of the SE relic from the centre of the cluster is $\gtrsim 1$~Mpc. For a typical shock speed of $\sim 1000$~${\rm km\,\,s^{-1}}$, the shock would have needed over $\sim 1$~Gyr to travel from the centre of the cluster to its current position. However, optical observations with the \emph{Telescopio Nazionale Galileo} revealed at least three galaxy subclumps: a strong galaxy peak near the SE radio relic, a more central galaxy peak, and another galaxy peak near the NW radio relic \citep{Boschin2013}. The NW and SE galaxy peaks appear to correspond to substructures moving apart from each other, and their movement could have recently triggered young radio relics in the ICM. 

The flat spectral indices could, in principle, also be explained if the merger is seen close to face-on, rather than in the plane of the sky. \citet{Skillman2013} suggest that for relics observed face-on, the integrated spectral index measured in the radio will actually be closer to the injection spectral index. However, a face-on orientation seems rather unlikely in ZwCl~2341.1+0000, because there is a distance of $\sim 2$~Mpc separating the NW and SE shocks/relics. Moreover, the brightness of the relics in a face-on orientation would probably be too low for the relics to be visible in the radio \citep{Vazza2011}.

We note, however, that caution is required in interpreting the spectral indices calculated by \citet{vanWeeren2009b} and by \citet{Giovannini2010}. The results of \citet{vanWeeren2009b} are based on \vla\ data at 325 and 1400 MHz \citep{Bagchi2002} and \gmrt\ data at 157, 241, and 610 MHz, and the spectral indices calculated from the \vla\ and \gmrt\ flux measurements do not take into account the different uv-coverages and sensitivies of the telescopes. Differences in sensitivities and uv-coverages could also affect the results presented by \citet{Giovannini2010}. Higher quality radio data are required to draw firmer conclusions about the relation between the spectral indices of the radio relics and the shock compression factors measured in X-ray.


\subsection{No shock across part of the SE relic?}

\begin{figure}
 	\begin{center}
		\includegraphics[width=\columnwidth,keepaspectratio=true,clip=true,trim=0mm 65mm 35mm 0mm]{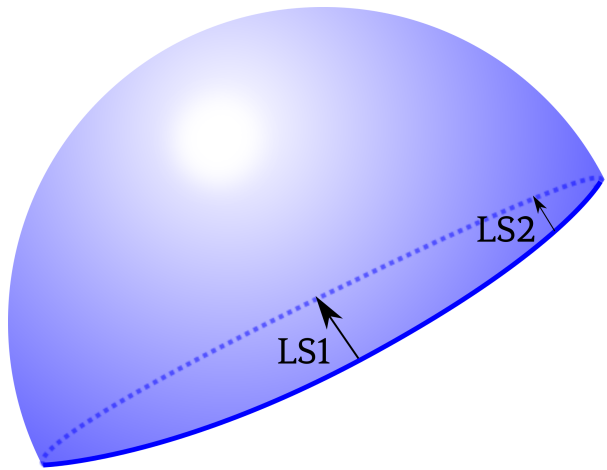}
 	\end{center}
	\caption{Drawing of a possible geometry in which the brightness jump across the edges of the radio relic might be to faint for the shock to be detected in the archival X-ray observations of ZwCl~2341.1+0000. The shock is shown as a spherical cap. The line of sight through the centre of the spherical cap (LS1) is shorter than the line of sight through the edges (LS2), which will cause the surface brightness to decrease towards the edges. However, the same trend should be seen in the brightness of the radio relic, and this is not observed at the SE relic in ZwCl~2341.1+0000.}
	 \label{fig:geometry3}
\end{figure}

If radio relics are formed by particle (re-)acceleration at shock fronts, then a shock should span the whole length of the radio relic it traces. Yet, this expectation is at odds with the results in ZwCl~2341.1+0000, in which the SE shock was found to subtend an arc that is only about a third of the length of the arc subtended by the SE radio relic; the result is confirmed with a confidence of $\sim 90\%$.

Because the shock is observed only along the very central part of the radio relic, it is unlikely that the lack of a shock across both the NE and the SW parts of the SE relic is caused by projection effects. It would be possible for the shock to be missed if across the ``ends'' of the radio relic the shock's distance from the cluster centre was larger than across the relic's central part, since the radius of the discontinuity was kept fixed in the fits (a discontinuity at a smaller radius was excluded by eye). This would imply that the shock has a convex shape with respect to the cluster centre; such a shock morphology has not been seen in numerical simulations, nor (yet) in other observations.

Another possibility is that the shock is more difficult to detect at the edges of the radio relic because of a shorter line of sight through the shock surface. For a spherical cap shock surface, the line of sight through the edges of the cap will be shorter than through the centre, which could cause the shock to gradually get fainter in X-ray from the centre to the edges of the cap, to the point where it escapes detection (see Fig. \ref{fig:geometry3}). In this geometry, the radio emission would follow the same trend. However, at the SE relic in ZwCl~2341.1+0000, the radio brightness in regions without a shock, especially in the SW part of the relic, is similar to the radio brightness in the region in which a shock is detected in X-ray.

As we discussed in the previous subsection, if the geometry deviates from that of a prolate spheroid and the cluster is in fact more extended along the line of sight, then the density compression factors measured at the putative shock fronts would be even lower. This would strengthen our conclusion that no density discontinuity is present along part of the SE radio relic.

Only deeper X-ray data could reveal the elusive shock across parts of the SE radio relic. However, if deeper observations confirm the results derived from the archival \xmm\ and \chandra\ data, and there is indeed no shock across the ends of the SE relic, then this would pose a significant challenge to our current theories about the formation of radio relics.

\section*{Acknowledgments}

We thank the referee for helpful comments. GAO is supported by the research group SFB~676, funded by the Deutsche Forschungsgemeinschaft (DFG). MB acknowledges support by the research group FOR~1254, funded by the DFG. RJvW acknowledges support provided by NASA through Einstein Postdoctoral Fellowship grant number PF2-130104 awarded by the Chandra X-ray Center, which is operated by the Smithsonian Astrophysical Observatory for NASA under contract NAS8-03060. This research is based on data from observations obtained with \xmm, an ESA science mission with instruments and contributions directly funded by ESA Member States and the USA (NASA). This research has also made use of data obtained from the \chandra\ Data Archive, and of software provided by the \chandra\ X-ray Center (CXC) in the application packages {\sc ciao}, {\sc ChIPS}, and {\sc Sherpa}. 

\bibliographystyle{mn2e}
\bibliography{bibliography}

\label{lastpage}

\end{document}